\documentclass[oldversion]{aa}
\usepackage{graphicx}
\usepackage{txfonts}
\usepackage{aalongtable}
\usepackage{multirow}

\newcommand{\teff}{T_\mathrm{eff}}
\newcommand{\logg}{\log g}
\newcommand{\feh}{\mathrm{[Fe/H]}}
\newcommand{\ofe}{\mathrm{[O/Fe]}}
\newcommand{\oxh}{\mathrm{[O/H]}}

\begin{document}

\title{Oxygen Abundances in Nearby Stars
\thanks{Tables 4 to 6 are only available in electronic form at the CDS via anonymous ftp to {\tt cdsarc.u-strasbg.fr (130.79.128.5)} or via {\tt http://cdsweb.u-strasbg.fr/cgi-bin/qcat?J/A+A/}} \fnmsep
\thanks{Partially based on observations obtained with the Hobby-Eberly Telescope, which is a joint project of the University of Texas at Austin, the Pennsylvania State University, Stanford University, Ludwig-Maximilians-Universit\"at M\"unchen, and Georg-August-Universit\"at G\"ottingen; and data from the UVES Paranal Observatory Project (ESO~DDT~Program ID~266.D-5655).}
       }
\subtitle{Clues to the formation and evolution of the Galactic disk}

\author{I. Ram\'irez, C. Allende~Prieto, \and D. L. Lambert}

\authorrunning{Ram\'irez et~al.}

\institute{McDonald Observatory and Department of Astronomy, University of Texas at Austin, RLM 15.306 Austin, TX, 78712-1083 \\ \email{[ivan;callende;dll]@astro.as.utexas.edu}}

%\date{A\&A, in press}
%\date{Submitted to A\&A, October 22; December 7 (revised)}
\date{Received October 22; accepted December 12}

\abstract{The abundances of iron and oxygen are homogeneously determined in a sample of 523 nearby ($d<150$~pc) FGK disk and halo stars with metallicities in the range $-1.5<\feh<0.5$. Iron abundances were obtained from an LTE analysis of a large set of \ion{Fe}{i} and \ion{Fe}{ii} lines with reliable atomic data. Oxygen abundances were inferred from a restricted non-LTE analysis of the 777~nm \ion{O}{i} triplet. We adopted the infrared flux method temperature scale and surface gravities based on \textit{Hipparcos} trigonometric parallaxes. Within this framework, the ionization balance of iron lines is not satisfied: the mean abundances from the \ion{Fe}{i} lines are systematically lower by 0.06~dex than those from the \ion{Fe}{ii} lines for dwarf stars of $\teff>5500$~K and $\feh<0.0$, and giant stars of all temperatures and metallicities covered by our sample. The discrepancy worsens for cooler and metal-rich main-sequence stars. We use the stellar kinematics to compute the probabilities of our sample stars to be members of the thin disk, thick disk, or halo of the Galaxy. We find that the majority of the kinematically-selected thick-disk stars show larger [O/Fe] ratios compared to thin-disk stars while the rest show thin-disk abundances, which suggests that the latter are thin-disk members with unusual (hotter) kinematics. A close examination of this pattern for disk stars with ambiguous probabilities shows that an intermediate population with properties between those of the thin and thick disks does not exist, at least in the solar neighborhood. Excluding the stars with unusual kinematics, we find that thick-disk stars show slowly decreasing $\ofe$ ratios from about 0.5 to 0.4 in the $-0.8<\feh<-0.3$ range. Using a simple model for the chemical evolution of the thick disk we show that this trend results directly from the metallicity dependence of the Type~II supernova yields. At $\feh>-0.3$, we find no obvious indication of a sudden decrease (i.e., a `knee') in the $\ofe$ vs. $\feh$ pattern of thick-disk stars that would connect the thick and thin disk trends at a high metallicity. We conclude that Type~Ia supernovae (SN~Ia) did not contribute significantly to the chemical enrichment of the thick disk. In the $-0.8<\feh<+0.3$ range, thin-disk stars show decreasing $\ofe$ ratios from about 0.4 to 0.0 that require a SN~Ia contribution. The implications of these results for studies of the formation and evolution of the Galactic disk are discussed.
}

\maketitle

\section{Introduction}

Nearby stars are ideal targets for chemical abundance studies due to the availability of accurate trigonometric parallaxes, a negligible reddening, and their brightness, which allows us to estimate the stellar atmospheric parameters as well as acquire high resolution, high signal-to-noise spectra of these objects.

With the Sun practically in the Galactic plane, the solar neighborhood contains mainly thin-disk objects, with a still significant fraction of thick-disk stars and only a few halo members. Distinctive patterns in the chemical composition of certain groups of stars contain valuable information for studies of Galaxy formation, evolution, the chemical enrichment history of the interstellar medium, and the nature of the nucleosynthesis sites. This paper deals with the determination of iron and oxygen abundances in nearby stars.

Due to the large number of lines and the availability of reliable atomic data, iron is an excellent reference element to assess the overall chemical composition of FGK stars. In fact, the term ``metallicity,'' denoted as $\feh$, usually refers to the iron abundance only.\footnote{Throughout this paper, we use the standard definitions $\mathrm[X/Y]=\log(N_X/N_Y)-\log(N_X/N_Y)_\odot$ and $A_\mathrm{X}=\log(N_X/N_H)+12$, where $N_X$ is the number density of the element $X$.} Oxygen is also one of the most important elements in chemical abundance studies. It is the third more abundant element in the Universe, after hydrogen and helium, and its abundance has a relatively important effect on the internal structure and evolution of stars. 

Unfortunately, the determination of stellar abundances is not straightforward. Since they are the basis upon which many astrophysical studies rest, it is important to examine carefully the methods of abundance determination and possible errors associated with them. Oxygen is a particularly difficult element, with only a few lines available to analyze and with all of them subject to different model complications and/or observational difficulties. For example, in main-sequence FGK stars, the forbidden [\ion{O}{i}] lines at 630~nm and 636~nm are both very weak and the 630~nm line is blended with a Ni line; the UV and IR OH lines are both very difficult to observe, the former due to atmospheric extinction and low CCD sensitivity and the latter because they are weak lines. The UV~OH lines are also strongly affected by surface inhomogeneities (Asplund \& Garc\'ia P\'erez 2001). Finally, the \ion{O}{i} lines at 777 nm, which we use as our oxygen abundance indicators, even though easy to observe, are severely affected by departures from LTE (e.g., Kiselman 1993).

\bigskip

The global structure of the Galaxy has been revealed by analyses of star counts. In most of these studies, the star count data are well approximated by two double-exponential disks, for the thin and thick disks, and two spheroids, for the bulge and halo (e.g., Chen et~al. 2001, Siegel et~al. 2002, Larsen \& Humphreys 2003, Robin et~al. 2003, Juri\'c et~al. 2005).\footnote{In general, the double-exponential disks have the form: $\rho(R,z)=\rho_\odot e^{-(R-R_\odot)/h_R} e^{-\mid z\mid/h_z}$, where $R$ is the galactocentric distance, $z$ the height above the Galactic plane, $\rho_\odot$ and $R_\odot$ the local values of $\rho$ and $R$, $h_R$ the radial scale length, and $h_z$ the vertical scale length. As for the spheroids, there is a wide variety of density laws (see Sect.~6.1 in Siegel et~al. 2002).} Using star counts, Gilmore \& Reid (1983) were the first to provide compelling evidence for the existence of the Galactic thick disk. They suggested vertical scale lengths of 300~pc for the thin disk and 1450~pc for the thick disk, as well as a 2\% relative density of thick-disk stars in the solar neighborhood. More recent studies have confirmed the existence of the two disks in star counts, although the values of the scale lengths (both vertical and radial) as well as the local relative density of thick-disk stars are still under debate (Chen et~al. 2001, Siegel et~al. 2002, Cabrera-Lavers et~al. 2005). Values reported in the literature for the vertical scale length range from 100 to 400 pc for the thin disk and from 600 to 1500~pc for the thick disk. The radial scale length of the thick disk ($\sim3$ kpc) also seems to be larger than that of the thin disk ($\sim2$ kpc), and estimates of the local relative density of thick disk stars range from 2 to 15\%. Thick disks have also been observed in many edge-on spiral galaxies, where bright thin disks are seen surrounded by faint red thick disks in deep photometry (e.g., Dalcanton \& Bernstein 2002). 

The thin- and thick-disk stars of our Galaxy have distinct Galactic space velocities: thick-disk stars appear to be rotating slower around the center of the Galaxy (mean drift velocity relative to the Sun of about $-50$~km~s$^{-1}$) and have wider velocity dispersions compared to thin-disk stars (mean drift velocity of about $-10$~km~s$^{-1}$; e.g., Soubiran 1993, Soubiran et~al. 2003).

Chemical abundance surveys demonstrate the metal-poor nature of the thick disk compared to the local thin disk (e.g., Gilmore et~al. 1995, Allende~Prieto et~al. 2006). The $\feh$ distribution of the thick disk spans the range $-1.0<\feh<-0.3$ while that of the thin disk is $-0.8<\feh<+0.4$.\footnote{In this paper, the so-called metal-weak thick-disk is not studied due to the low number of stars with $\feh<-1$ in the sample; the reader is referred to Beers et~at. (2002) for a discussion of this topic.} It seems that no thin-disk stars with $\feh<-0.8$ exist whereas the existence of thick-disk stars with $\feh>-0.3$ is under debate.

Fuhrmann (1998) has suggested an age gap, at about 9--10~Gyr, between thin- and thick-disk stars, such that all thick-disk stars are older than thin-disk stars. Later studies (e.g., Gratton et~al. 2000, Bensby et~al. 2005, Reddy et~al. 2006, Allende~Prieto et~al. 2006) have confirmed the old nature of thick disk stars but the age gap is difficult to detect due to uncertainties and systematic errors in the stellar age determination. Note, for example, that in their study of a very large sample of stars from the third data release of the \textit{Sloan Digital Sky Survey}, Allende~Prieto et~al. (2006) find thick-disk stars as young as 8~Gyr.

Regarding oxygen abundance patterns, they have been suggested to be distinct for thin- and thick-disk stars (e.g., Prochaska et~al. 2000, Bensby et~al. 2004): thick-disk stars appear to have larger [O/Fe] and [O/H] compared to thin disk stars at any given metallicity within the $\feh$ range in common. This implies that thin- and thick-disk stars were formed from two different mixtures of gas, presumably at different epochs and/or locations. Oxygen abundance patterns and their systematic differences between thin- and thick-disk stars thus contain important information about the formation and evolution of the Galactic disk. Other elements, but not all of them, also show distinct abundance patterns for thin- and thick- disk stars (e.g., Bensby et~al. 2005, Reddy et~al. 2006).

\bigskip

This paper is distributed as follows: basic data are described in Sect.~\ref{s:sample}, the stellar kinematics and the criterion adopted for thin disk, thick disk, or halo membership are discussed in Sect.~\ref{s:kinematics}, the abundance analysis is explained in Sect.~\ref{s:abundance} with the exception of the non-LTE corrections to the oxygen abundances, which are discussed separately in Sect.~\ref{s:nlte}. The oxygen vs. iron abundance trends are traced in Sect.~\ref{s:ofetrends}, stellar ages are discussed in Sect.~\ref{s:ages}, and interpretations of the results are given in Sect.~\ref{s:discussion}.

\section{Sample and Data Sources} \label{s:sample}

\subsection{Spectra} \label{s:spectra}

Most of the spectra used in this work were available from previous and ongoing studies of nearby stars and were thus already reduced using standard methods (see the references below for details). In addition, we obtained spectra of 22 stars to complement the available data. A summary of our data sources is given in Table~\ref{t:datasources}.

Spectra of about 100 stars within 15 pc from the Sun were available from the \textit{Spectroscopic Survey of Stars in the Solar Neighborhood} (S$^4$N) of Allende~Prieto et~al. (2004a, hereafter AP04a).\footnote{{\tt http://hebe.as.utexas.edu/s4n, http://www.astro.uu.se/$\sim$s4n}} Their observations were made with the 2.7~m Telescope at McDonald Observatory and the ESO 1.5~m Telescope at La Silla. Spectra of about 50 stars obtained with the 9.2~m Hobby Eberly Telescope (HET) as part of ongoing studies of the closest stars, metal-rich dwarfs and giants were also available for this work. We also used 44 FGK stars spectra from the online UVES-VLT library by Bagnulo et~al. (2003),\footnote{{\tt http:/www.eso.org/uvespop}} who took advantage of the availability of the UVES spectrograph on the VLT during part of twilight time to observe a significant number of bright stars covering most of the HR diagram as well as stars in two open clusters.

The major contribution of data to this work comes from the Reddy et~al. (2003, 2006; hereafter R03, R06) studies. About 180 of these spectra were used in the R03 chemical abundance work that consists mainly of thin-disk objects, while spectra of about 170 stars with halo and thick-disk kinematics were kindly provided by B.~Reddy prior to publication (R06).

Due to the lack of metal-rich stars with thick-disk kinematics in the samples described above, we obtained spectra of 22 stars with these characteristics. We used the 2dcoud\'e spectrograph on the Harlan J. Smith 2.7-m Telescope at McDonald Observatory (Tull et~al. 1995) with a resolving power of about 60,000. This is the same instrument used for the Reddy et~al. studies. Overscan subtraction, flat-fielding, scattered light removal, extraction, wavelength calibration, and continuum normalization were applied to the spectra using IRAF.\footnote{IRAF is distributed by the National Optical Astronomy Observatories, which are operated by the Association of Universities for Research in Astronomy, Inc., under cooperative agreement with the National Science Foundation -- {\tt http://iraf.noao.edu}}

Combining all data sets we ended up with 564 spectra for 523 stars, after discarding high rotational velocity stars, double-lined spectroscopic binaries, and very low S/N spectra. All spectra we used have high-resolution ($R\sim45,000-120,000$), moderate to high signal-to-noise ratios ($S/N\sim100-600$), and cover at least most of the wavelength range required for this work (4500~\AA\ -- 7800~\AA). For the stars that have spectra from more than one source, due to different values of the resolution, instead of combining the spectra, the analysis was done independently and the final results averaged.

To perform a differential abundance analysis, a solar spectrum was adopted from the S$^4$N database. This spectrum was obtained by co-adding skylight spectra taken when the angle between the Sun and the port that allows skylight to enter the slit room (directed towards the zenith) was between 35 and 60 degrees, which minimizes the distortion of the solar integrated spectrum due to scattering in the Earth's atmosphere (see, e.g., Gray et~al. 2000). A comparison of this spectrum with the solar flux atlas of Kurucz et~al. (1984), smoothed to match the resolution of the former, showed no evidence of dispersed light and that the observational error is instead dominated by the continuum placement (Sect.~2 in AP04a). In addition, the reflection spectrum of the asteroid Iris, used by R03, as well as another skylight spectrum from our latest observations, were used.

\begin{table}
\centering
\begin{tabular}{lcccr} \hline\hline
Source & Facility/Instrument & $R$ & $S/N$ & Spectra \\ \hline
\multirow{2}{*}{AP04a} & McD/2dcoud\'e & 60,000 & \multirow{2}{*}{150-600} & \multirow{2}{*}{102} \\
& ESO/FEROS & 45,000 & & \\
TW & HET/HRS & 120,000 & $\sim300$ & 49 \\
B03 & VLT/UVES & 80,000 & 300-500 & 44 \\
R03 & McD/2dcoud\'e & 60,000 & $\sim400$ & 178 \\
R06 & McD/2dcoud\'e & 60,000 & 100-200 & 169 \\
TW & McD/2dcoud\'e & 60,000 & $\sim250$ & 22 \\
\hline
\end{tabular}
\caption{Sources of data (spectra only) used in this work: AP04a (Allende Prieto et~al. 2004a), B03 (Bagnulo et~al. 2003), R03 (Reddy et~al. 2003), R06 (Reddy et~al. 2006), TW (this work). Facilities: McD (2.7~m Telescope at McDonald Observatory), ESO (1.5~m Telescope on La Silla), HET (Hobby-Eberly Telescope), VLT (Very Large Telescope).}
\label{t:datasources}
\end{table}

\subsection{Photometry}

Stellar magnitudes and colors were required to obtain surface gravities and effective temperatures from the location of the stars in the HR diagram and temperature-color calibrations, respectively (Sect.~\ref{s:parameters}). The General Catalogue of Photometric Data (Mermilliod et~al. 1997) provided magnitudes and colors of our sample stars in several systems (\textit{UBV}, $uvby$, Vilnius, Geneva, $RI_\mathrm{(C)}$, and DDO). Photometric data from the \textit{Hipparcos-Tycho} mission (ESA 1997) and the \textit{Two Micron All Sky Survey} (2MASS, Cutri et~al. 2003) were also used.

\begin{figure}
 \includegraphics[width=8.6cm]{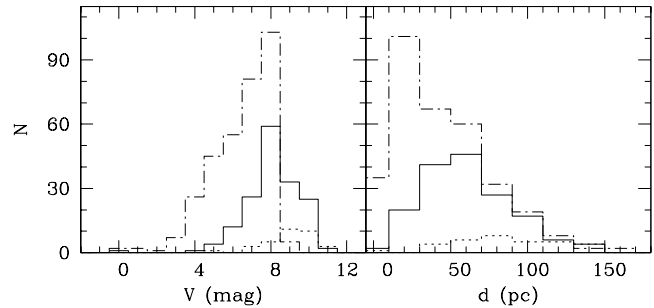}
 \caption{Distribution of our sample stars in $V$ magnitude and distance ($d$). Dash-dotted lines: $P_1>0.5$ (i.e. mostly thin-disk stars), solid lines: $P_2>0.5$ (thick-disk stars), dotted lines: $P_3>0.5$ (halo stars).}
 \label{f:vmagd}
\end{figure}

All our sample stars are in the \textit{Hipparcos} catalog and have parallaxes measured with uncertainties typically smaller than 10\% (the average of these parallax errors is 5.8\%). The distribution of our sample stars in $V$ magnitude and distance is given in Fig.~\ref{f:vmagd}. In this Figure, the stars have been assigned to either the thin disk, thick disk, or halo of the Galaxy, as described in Sect.~\ref{s:kinematic-criterion}. Most of our sample stars are brighter than $V\sim10$~mag and closer than $d\sim100$~pc. Note, however, that in order to collect a significant fraction of halo and thick-disk stars, a larger volume needs to be sampled and this is why, in general, our halo and thick-disk objects are fainter and more distant than the bulk of our thin-disk members.

In Fig.~\ref{f:vmagd} we see that most (91\%) of our stars are within 100~pc from the Sun. Therefore, for most of them, reddening is not an issue. For the farthest stars, however, corrections of the order of $0.01-0.02$ mag may be required if they are close to the Galactic plane, but they were not applied here. There is virtually no interstellar absorption within 75~pc from the Sun (see, e.g., Fig.~3 in Lallement et~al. 2003). This may also be the case for larger distances as the shape of the `Local Bubble' (a region of very low interstellar gas density, devoid of dust, around the Sun) is far from spherical (e.g., Leroy 1999) and stars as distant as 150 pc may not need a reddening correction, depending on their line of sight.

Using a combination of interstellar reddening maps and empirical laws, as described in Ram\'irez \& Mel\'endez (2005a, their Sect.~3.1), we estimate a reddening correction $E(B-V)\simeq0.01$~mag for stars with $d\simeq100$~pc located in the Galactic plane. For $d\simeq125$~pc in the Galactic plane, $E(B-V)\simeq0.02$. If considered, these reddening estimates of 0.01 and 0.02 will lead to higher effective temperatures by about 50~K and 100~K, respectively, at $\teff\simeq6000$~K. R06 applied reddening corrections to stars farther than 100~pc and found the maximum $E(B-V)$ to be about 0.03 (roughly equivalent to their quoted $A_V\sim0.1$), which corresponds to a maximum increase in $\teff$ of about 140~K. Note from Fig.~\ref{f:vmagd} that our sample contains a comparable number of thin- and thick-disk stars beyond 100~pc which implies that the errors introduced by neglecting reddening affect in a similar way our thin- and thick-disk subsamples.

\section{Kinematics} \label{s:kinematics}

\subsection{Radial velocities}

For most (94.1\%) of our sample stars, we adopted radial velocities ($V_r$) from the literature. A careful inspection of the values given by different authors allowed us to discard a few published results and/or assign realistic error bars to the literature data.

For 75.9\% of the sample we found $V_r$ values in the Nordstr\"om et~al. (2004) catalog and adopted them. Then, the radial velocitites given by AP04a and R06 were adopted for another 11.9\% of the sample stars that are not included in the Nordstr\"om et~al. catalog. Note that AP04a adopted mean values of literature data while R06 used published results and, for the stars without a previous determination of $V_r$, derived radial velocities by measuring wavelength shifts for a large set of spectral lines. For 6.3\% of our sample stars, which are not included in any of the studies cited above, we adopted the mean $V_r$ from the following sources, when available (we checked and avoided duplicity of original results among catalogs when computing the mean): Duflot et~al. (1995), Barbier-Brossat \& Figon (2000), Malaroda et~al. (2001), Latham et~al. (2002), and Valenti \& Fischer (2005). Finally, for only 31 stars (5.9\% of the sample) for which we did not find published $V_r$ values, we estimated radial velocities as follows: first, the central wavelength of 10 strong lines within the spectral range from 6000~\AA\ to 8000~\AA\ were measured and compared with those measured in the solar spectrum of Kurucz et~al. (1984); the wavelength shifts were then transformed into velocities that include the Earth's motion around the Sun; and finally, these were put into the heliocentric frame by computing heliocentric velocity corrections using the {\tt rvcorrect} task in IRAF.

The mean error in our adopted $V_r$ values is 0.8~km~s$^{-1}$ which reflects the accuracy of the literature data. For 91\% of our sample stars the error is smaller than 1~km~s$^{-1}$ and it is smaller than 5~km~s$^{-1}$ for 97.7\% of them. Only 5 stars have errors in $V_r$ that exceed 10~km~s$^{-1}$. These are stars with discrepant literature data, most likely single-lined binaries for which no orbit has been determined yet.

We used the radial velocities to compute Galactic space velocities and determine the membership probabilities of our sample stars to be thin-disk, thick-disk, or halo members. The accuracy of our adopted $V_r$ values is sufficient for the present purposes (see below).

\subsection{Galactic space velocities}

Using \textit{Hipparcos} data (coordinates, parallaxes, and proper motions) and the radial velocities obtained as described above we calculated the Galactic space velocities $U,V,W$ of our sample stars. These velocities are measured with respect to the Sun. The $V$ component is measured in the direction of Galactic rotation, $W$ is the component perpendicular to the plane of the Galaxy (positive towards the Galactic north pole), and $U$ is measured towards the Galactic center.

The transformation equations from radial velocity, parallax, and proper motion to Galactic space velocity we used are given by, e.g., Johnson \& Soderblom (1987). Note, however, that due to correlations between astrometric parameters, as given in the \textit{Hipparcos} catalog, the error estimates in this procedure need to be corrected (see the \textit{Hipparcos} catalog or the Appendix B in AP04a).

The typical error in the \textit{Hipparcos} proper motion is about 1~mas. Thus, the error in the heliocentric tangential velocities ($V_t$) of our sample stars due to this error is very small. It is only about 0.05~km~s$^{-1}$ for a star at 10~pc and, in our sample, it reaches a maximum of 0.8~km~s$^{-1}$ for a distance of 150~pc. Therefore, the error in $V_t$ is dominated by the error in the parallax. For example, for the stars in our sample for which the error in the parallax is about 1\%, the error in $V_t$ is about 0.5~km~s$^{-1}$ while for the stars with a 10\% error in the parallax the error in $V_t$ is about 10~km~s$^{-1}$. Examining our sample results, we find that for thick-disk (thin-disk) stars, the mean error in $V_t$ is about 8.4~(1.6)~km~s$^{-1}$. Mean errors in $U,V,W$ in our sample are 3.1, 4.0, and 2.5~km~s$^{-1}$, respectively. About 98~(91)~\% of our sample stars have uncertainties in $V_r$ smaller than 5~(1)~km~s$^{-1}$ which implies that the errors in $V_t$ dominate the uncertainties in the $U,V,W$ velocities. For the nearest stars, the error in $V_r$ is normally comparable with the uncertainty introduced by the \textit{Hipparcos} data, and only in a few extreme cases where the error in $V_r$ exceeds 10~km~s$^{-1}$, the errors in $U,V,W$ are dominated by the uncertainty in $V_r$.

\subsection{Kinematic criterion for subsampling} \label{s:kinematic-criterion}

Studies of stellar kinematics show that the distribution of $U,V,W$ velocities may be reasonably well decomposed by assuming that there are three stellar populations involved, each with a Gaussian velocity distribution: thin disk, thick disk, and halo (e.g., Soubiran 1993, Soubiran et~al. 2003). As a whole, the thick disk rotates slower around the center of the Galaxy compared to the thin disk and it has a wider velocity dispersion. The halo, on the other hand, has little or no overall rotation but the widest velocity dispersion.

Typical values for the rotational lags with respect to the Sun are $V\sim-10$~km~s$^{-1}$ for the thin disk, $V\sim-50$~km~s$^{-1}$ for the thick disk, and $V\sim-200$~km~s$^{-1}$ for the halo. There is still controversy regarding the exact values of both the rotational lags and the velocity dispersions but it is clear that the absolute value of the rotational lag and the velocity dispersions for the halo and thick disk are larger than those of the thin disk (compare, e.g., the value for the rotational lag of the thick disk by Nissen 1995, $\sim-$50~km~s$^{-1}$, with that given by Chiba \& Beers 2000, $\sim-$30~km~s$^{-1}$, and the relatively large velocity dispersions for the thin disk according to Soubiran et~al. 2003 with those given by AP04a).\footnote{However, it is possible that these differences are due to the use of samples with different characteristics. Note, for example, that the $V$ velocities depend on spectral type (e.g., Dehnen \& Binney 1998).} The mean thin/thick disk kinematics we adopted are given in Table \ref{t:parameterization}, and correspond to those inferred by Soubiran et~al. (2003). For the halo kinematics we adopted the values derived by Chiba \& Beers (2000) for their `pure' halo component. Note that the values in Table~\ref{t:parameterization} are given in the heliocentric frame.

\begin{table}
\centering
\begin{tabular}{lccccc} \hline\hline
 & $p_i$ & $V$ & $\sigma_U$ & $\sigma_V$ & $\sigma_W$ \\ \hline
 i=1 , thin disk  & 0.90 & $ -12$ &  39 &  20 & 20 \\
 i=2 , thick disk & 0.08 & $ -51$ &  63 &  39 & 39 \\ 
 i=3 , halo       & 0.02 & $-199$ & 141 & 106 & 94 \\ 
\hline
\end{tabular}
\caption{Parameterization adopted for the criterion of thin-disk, thick-disk, or halo membership. Units are km~s$^{-1}$, except for $p_i$, which is dimensionless.}
\label{t:parameterization}
\end{table}

\begin{figure*}
 \centering
 \includegraphics[width=15cm]{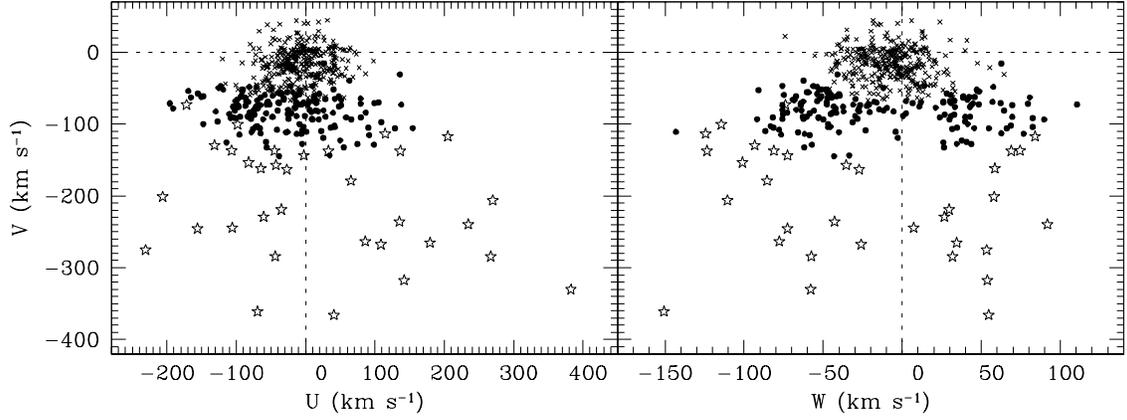}
 \caption{Distribution of our sample stars in $(U,V,W)$ Galactic space velocities. Crosses: $P_1>0.5$ (thin-disk stars), circles: $P_2>0.5$ (thick-disk stars), stars: $P_3>0.5$ (halo stars). The dotted lines intersect at $U=V=W=0$.}
 \label{f:xyzuvw}
\end{figure*}

Following the formulation adopted in previous studies (e.g., Mishenina et~al. 2004, Bensby et~al. 2005, R06), the probability that a star belongs to the component that has a mean $V=V_i$ and a velocity dispersion $\sigma_V=\sigma_{V_i}$, based only on the $V$ velocity of the star, is proportional to $\exp\,[-(V-V_i)^2/(2\sigma_V^2)]$. Combining the three velocities, the probability of a star to belong to one of the components is
\[
P_i=\frac{c_i}{(2\pi)^{3/2}\sigma_{U_i}\sigma_{V_i}\sigma_{W_i}}\exp{ \left\{ -0.5 \left[\frac{U^2}{\sigma_{U_i}^2}+\frac{(V-V_i)^2}{\sigma_{V_i}^2}+\frac{W^2}{\sigma_{W_i}^2}\right] \right\} } \ ,
\]
where $i=1$ for the thin disk, $i=2$ for the thick disk, $i=3$ for the halo, and $c_i$ is a normalization constant given by
\[
c_i=\frac{p_i}{\sum_{i=1}^3 p_i(P_i/c_i)}\ .
\]
In the last equation, the $p_i$ values are the relative number densities of thin-disk, thick-disk, and halo stars in the solar neighborhood. The equations are given in this form because the ratio $P_i/c_i$ can be computed directly from the observed data.

The $p_i$ values are poorly known; values reported in the literature for $p_2$ (the relative density of thick-disk stars), for example, range between 0.02 and 0.15 (e.g., in order of increasing $p_2$: Reid \& Majewski 1993, Ojha 2001, Reyl\'e \& Robin 2001, Soubiran et~al. 2003). Here we adopted $p_1=0.90$, $p_2=0.08$, $p_3=0.02$. As revealed by star counts, the mean densities of thin-disk, thick-disk, and halo stars vary with distance from the Galactic plane ($z$). In a more general formulation, the $p_i$'s would need to be considered as functions of $z$. However, since our sample stars are all nearby, the $z$-dependence does not have a noticeable effect and the use of constant local $p_i$ values is valid.

\begin{figure}
 \includegraphics[width=8.6cm]{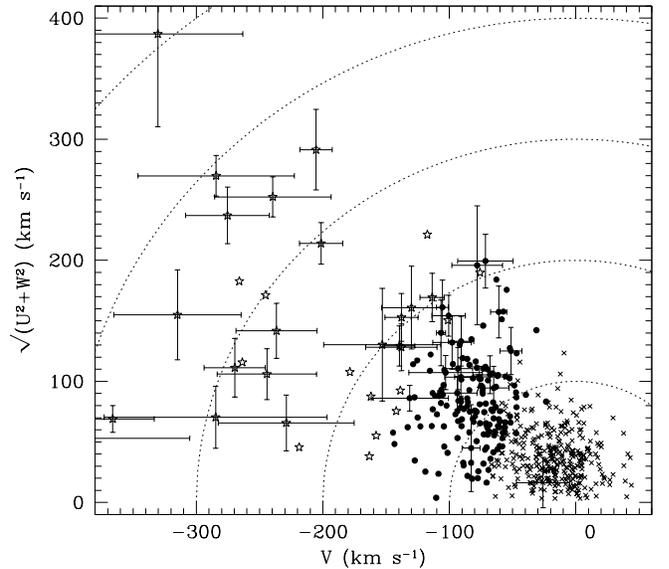}
 \caption{Toomre diagram for our sample stars. Symbols are the same as in Fig.~\ref{f:xyzuvw}. Error bars are shown only for stars with uncertainties larger than 20~km~s$^{-1}$ in $\sqrt{U^2+V^2+W^2}$.}
 \label{f:toomre}
\end{figure}

Figs.~\ref{f:xyzuvw} and \ref{f:toomre} illustrate the kinematic properties of our sample stars and the membership criterion adopted for subsampling. Table 4, available at the CDS, provides the basic data and kinematic properties of our sample stars, including the probabilities described above.

It is important to notice that the errors in the $P_i$ probabilities due to errors in the $U,V,W$ velocities are only important for stars in the `transition' regions of the Galactic velocity space. This is because even though there is an overlap in the kinematical distributions of the three groups, the peaks in $V$ are still well defined (see Figs.~\ref{f:xyzuvw}, \ref{f:toomre}, and Table~\ref{t:parameterization}) and a given error in the $U,V,W$ set leads to a larger error in $P_i$ if $P_i\sim0.5$ compared to the $P_i\sim1.0$ case. By propagating the errors in the $U,V,W$ velocities to the $P_i$ values of each star we find there are only 0, 3, and 1 thin-disk (with $P_1>0.7$), thick-disk ($P_2>0.7$), and halo ($P_3>0.7$) stars, respectively, for which the uncertainty in $P_i$ is larger than 0.2 (the largest error found was 0.26). The corresponding mean values are $<\sigma(P_1)>=0.005$, $<\sigma(P_2)>=0.043$, and $<\sigma(P_3)>=0.043$. Therefore, if we set a $P_i>0.7$ constrain to determine the membership of the sample stars to the $i$-group, the chance of having a star with $P_i<0.5$ after considering errors in the $U,V,W$ data is essentially zero.

\section{Abundance Analysis} \label{s:abundance}

Both the iron and oxygen abundances were determined by matching the observed line equivalent widths with those predicted by spectrum synthesis. A differential analysis was performed using the solar spectrum as a reference.

The classical 1D-LTE model atmospheres, without convective overshooting, by Kurucz (1993), interpolated to the ($\teff$\,,\,$\logg$\,,\,$\feh$) set of each sample star, were adopted.\footnote{These are the models labeled `nover' in the R.~L.~Kurucz's website: {\tt http://kurucz.harvard.edu/grids.html}. We did not use the more recent and improved `odfnew' models but we found that the changes in the abundances if we used them instead of the `nover' models are small (less than 0.02~dex in the case of iron) and in any case systematic so the relative abundances and abundance ratios are not affected.} Even though the solar model with convective overshooting reproduces better the solar spectrum, the derivation of stellar parameters seems to be more robust with no-overshoot models (Castelli et~al. 1997). Thus, we preferred the later type of Kurucz models for this work.

The latest version of the spectrum synthesis code MOOG (Sneden 1973)\footnote{{\tt http://verdi.as.utexas.edu/moog.html}} was used for the LTE calculations. Since the oxygen lines require a non-LTE treatment, the equations of statistical equilibrium were solved with TLUSTY and non-LTE spectrum synthesis was done with SYNSPEC (see Hubeny 1988, Hubeny \& Lanz 1995).\footnote{{\tt http://nova.astro.umd.edu}} A detailed discussion of the non-LTE calculations is given separately in Sect.~\ref{s:nlte}.

\subsection{Equivalent width measurements} \label{s:ew}

For almost all the spectral lines that we used, the Lorentzian wings could be neglected (or were avoided, since we did not use very strong lines affected by saturation) so the equivalent widths ($EW$) were measured by fitting Gaussian profiles to the spectral lines. The $EW$ measurements were made using an automated algorithm that finds the central wavelength ($\lambda$) of a line within a certain wavelength range and fits a Gaussian profile centered on this wavelength. Possible blends are taken into account in this procedure.

The estimated error in our automatic $EW$ measurements is 3.5\% (1.8~m\AA\ for a typical 50~m\AA\ line), based on a comparison of $EW$ measurements of \ion{Fe}{i} lines on the same star but using spectra from three of the data sources described in Sect.~\ref{s:spectra}. This comparison also showed no systematic differences in the $EW$s measured in spectra from different sources.

The equivalent widths measured in this way for the three available solar spectra (see Sect.~\ref{s:spectra}) agree at the 1\% level, with the small differences in $EW$ being random and not systematic. The adopted solar equivalent widths are a weighted mean of the $EW$s measured in each of the three spectra, with a higher weight given to the skylight spectrum from the S$^4$N database.

\subsection{Line data and solar analysis} \label{s:ldandsolar}

We selected 119 \ion{Fe}{i} and 13 \ion{Fe}{ii} lines to determine the $\feh$ values of our sample stars. All these lines have transition probabilities measured in the laboratory and were chosen such that no evident blends with other features are present in the solar spectrum. Our \ion{Fe}{i} line selection also required the final line list not to show any correlation between the excitation potential ($EP$) and the reduced equivalent width ($REW=EW/\lambda$) of the lines, which makes it ideal for purely spectroscopic studies, where these correlations lead to degenerate solutions for the stellar parameters that are estimated by making the line-by-line abundances independent of $EP$ and $REW$. Note, however, that our procedure is not purely spectroscopic.

We adopted laboratory $gf$-values for the \ion{Fe}{i} lines from different sources, all of which are described and compared to each other by Lambert et~al. (1996). When available, the Oxford values (e.g., Blackwell et~al. 1976) were adopted, otherwise a weighted mean with the typical uncertainties of the other sources was used. Transition probabilities for the \ion{Fe}{ii} lines are from the compilations of Allende~Prieto et~al. (2002) and Mel\'endez et~al. (2006). No astrophysical $gf$-values were used.

Oxygen abundances were determined only from the triplet \ion{O}{i} lines at 777~nm. The transition probabilities adopted are those given in the NIST database;\footnote{{\tt http://physics.nist.gov/PhysRefData/ASD/ \\ lines\_form.html}; see also Wiese et~al. (1996).} i.e., the average of the values calculated by Hibbert et~al. (1991), Butler \& Zeippen (1991), and Bi\'emont \& Zeippen (1992). In general, in order to analyze the [\ion{O}{i}] lines at 630~nm and 636~nm we would require higher resolution and S/N. The exceptions are most of the S$^4$N spectra, which are suitable for this type of analysis. However, we also avoided the 630~nm line due to the well-known blend with a \ion{Ni}{i} line (e.g., Allende~Prieto et~al. 2001), which requires an accurate determination of the Ni abundance prior to the oxygen abundance analysis.

Van der Waals damping constants were adopted mainly from the Barklem et~al. (2000) and Barklem \& Aspelund-Johansson (2005) calculations. For the few \ion{Fe}{i} lines that are not included in the Barklem et~al. (2000) work, the classical Uns\"old formula, with an enhancement factor of 2, was adopted. Standard approximations for the radiative (e.g., Gray 1992) and Stark (e.g., Cowley 1971) broadening constants, as coded in MOOG, were also adopted.

\begin{figure}
 \includegraphics[width=8.5cm]{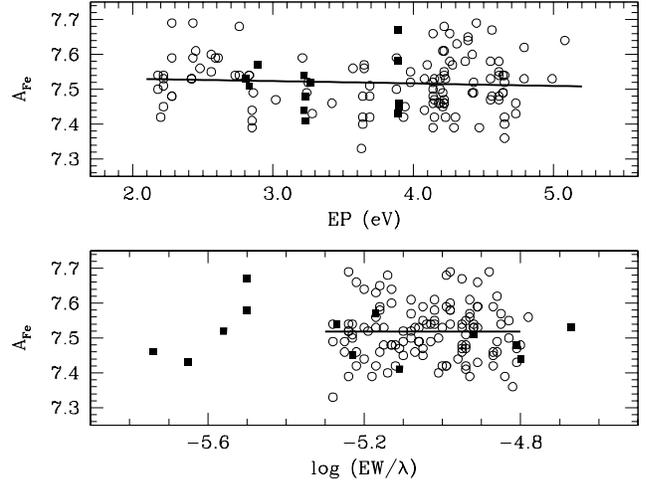}
 \caption{Our derived solar iron abundances from \ion{Fe}{i} (circles) and \ion{Fe}{ii} (squares) lines as a function of excitation potential and reduced equivalent width. Linear fits to the \ion{Fe}{i} data are also shown.}
 \label{f:sun}
\end{figure}

The line-list and atomic data adopted in this work are given in Table 5, available at the CDS.

\bigskip

For the Sun, we obtained $A_\mathrm{Fe}=7.52\pm0.08$ from the \ion{Fe}{i} lines, and $A_\mathrm{Fe}=7.51\pm0.07$ from the \ion{Fe}{ii} lines.\footnote{In this paper, all error bars given for the abundances are 1$\sigma$ errors, where $\sigma$ is the standard deviation of the line-by-line abundances.} A microturbulent velocity of 0.90~km~s$^{-1}$ was chosen to make the abundances from the \ion{Fe}{i} lines independent of $REW$. As shown in Fig.~\ref{f:sun}, no significant correlation with the $EP$ of the lines is present either.

Our mean solar iron abundance of $A_\mathrm{Fe}=7.5\pm0.1$ is consistent, within error bars, with most other determinations using both classical 1D-LTE models (e.g., Holweger et~al. 1991) and 3D hydrodynamical simulations (e.g., Shchukina \& Trujillo~Bueno 2001; Asplund et~al. 2004, 2005).

An LTE solar oxygen abundance of $A_\mathrm{O}=8.85$ is obtained from the triplet. When non-LTE corrected, as in Sect.~\ref{s:nlte}, this value reduces to $A_\mathrm{O}=8.72$, in good agreement with most recent determinations of the solar oxygen abundance (e.g., Allende~Prieto et~al. 2001; Asplund et~al. 2004, 2005; Mel\'endez 2004; Shchukina et~al. 2005). The agreement is perfect if we consider only the oxygen abundance derived from the triplet, corrected for non-LTE and using one-dimensional model atmospheres, but note that granulation effects reduce this abundance to 8.66 (see Asplund et~al. 2004).

\subsection{Atmospheric parameters} \label{s:parameters}

\subsubsection{Effective temperature} \label{s:teff}

The infrared flux method (IRFM) color-temperature calibrations by Ram\'irez \& Mel\'endez (2005b) were used to estimate the effective temperatures of our sample stars. They provide calibrations for 17 colors in different photometric systems. We used as many as possible of the 17 colors available for each star. On average, 8 colors were used, independently of the population (thin-disk, thick-disk, or halo).

The IRFM temperatures have been shown to be in good agreement with those derived from basic principles, i.e. with direct temperatures inferred from angular diameter and bolometric flux measurements (e.g., Ram\'irez \& Mel\'endez 2004, 2005a). When compared to temperatures spectroscopically derived from the excitation balance of \ion{Fe}{i} lines, however, a systematic difference of about 100~K is observed for metal-rich main sequence stars (e.g., Ram\'irez \& Mel\'endez 2004, Santos et~al. 2004, Yong et~al. 2004, Heiter \& Luck 2005). Interestingly, Santos et~al. (2005) have found reasonably good agreement between spectroscopic and direct temperatures for four of their sample stars while Casagrande et~al. (2006) have shown that the zero point of the IRFM $\teff$ scale can be in good agreement with the spectroscopic one depending on the flux calibration adopted. Masana et~al. (2006) have also found a difference of 100~K with the IRFM $\teff$ scale using their spectral energy distribution fit method. Besides this evidence for a systematic error in the zero point of the effective temperature scale, the use of several of the IRFM $\teff$ vs. color calibrations by Ram\'irez \& Mel\'endez (2005b) guarantees a very homogeneous, internally accurate, $\teff$ scale for our sample stars.

\subsubsection{Surface gravity}

The high accuracy of the \textit{Hipparcos} trigonometric parallaxes allowed us to determine reliable surface gravities for our sample stars. The procedure is outlined in R03 and AP04a.

An estimate of the stellar masses is needed to obtain the $\logg$ values. These were derived from a comparison of the location of the stars in the HR diagram with the theoretical isochrones by Bertelli et~al. (1994). This approach also provides age estimates (Sect.~\ref{s:ages}). Errors in the mass determined in this way may be as high as 10\%, considering not only errors in the fundamental stellar parameters but also systematic errors in the theoretical isochrones. Nevertheless, this error affects only marginally the $\logg$ value whereas the error in the parallax affects it substantially (see, e.g., Eq. 2 in Allende~Prieto et~al. 1999).

Given that the errors in the parallax of our sample stars are usually lower than 10\%, the typical error in $\logg$ is about 0.05~dex, which has only a marginal impact on the abundance determination.

An alternative approach to the $\logg$ value is to tune it until the mean Fe abundances obtained separately from \ion{Fe}{i} and \ion{Fe}{ii} lines agree, keeping all the other parameters constant. Allende~Prieto et~al. (1999) provide a comprehensive comparison of these spectroscopic surface gravities with those determined from \textit{Hipparcos} parallaxes. The spectroscopic procedure, however, is challenged by model uncertainties. Shchukina \& Trujillo Bueno (2001), for example, have shown that overionization due to near-UV radiation may lead to a \ion{Fe}{ii}/\ion{Fe}{i} ratio that differs from that obtained from the Saha equation in the solar photosphere. Similar results have been obtained by Th\'evenin \& Idiart (1999) and Korn et~al. (2003) but with a smaller \ion{Fe}{ii} vs. \ion{Fe}{i} difference. On the metal-poor side, Ram\'{\i}rez et~al. (2006) have demonstrated that the spectroscopic $\logg$ value of at least one star is in disagreement with the more reliable surface gravity obtained from fitting of the wings of strong lines.

\subsection{Abundances}

We determined absolute abundances from every spectral line of each star. Then, by comparing these results with the solar abundances, on a line-to-line basis, we inferred differential abundances for both iron and oxygen.

The atmospheric parameters and iron abundances were determined iteratively starting with $\feh=0$ models. After three iterations, neither $\teff$ nor $\logg$ showed further changes. At every iteration, a microturbulent velocity $v_t$ was determined by making the abundances from the \ion{Fe}{i} lines independent of $REW$.

\subsubsection{The ionization balance problem} \label{s:noionization}

\begin{figure}
 \includegraphics[width=8.5cm]{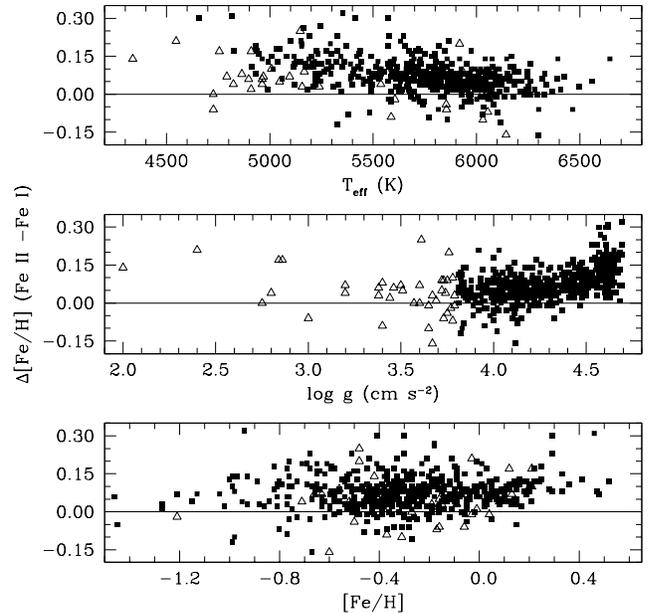}
 \caption{Difference between the $\feh$ values obtained with the \ion{Fe}{ii} and \ion{Fe}{i} lines as a function of atmospheric parameters. Giant stars ($\logg<3.8$) are shown with triangles and dwarf stars ($\logg>3.8$) with squares.}
 \label{f:fe1vsfe2}
\end{figure}

The differences between the $\feh$ values derived with \ion{Fe}{i} and \ion{Fe}{ii} lines are shown in Fig.~\ref{f:fe1vsfe2} as a function of atmospheric parameters. It is clear that, on average, for $\teff>5500$~K, $\logg<4.4$, and $\feh<0.1$, the \ion{Fe}{ii} abundances are systematically larger than the \ion{Fe}{i} abundances by 0.06 dex. The disagreement worsens for cooler, higher surface gravity, and more metal-rich dwarf stars. Cool giants with 4700~K$<\teff<5000$~K also show systematically large \ion{Fe}{ii} abundances but at a roughly constant level of about 0.06 dex. The solution to this problem is at present unknown, although it could be explained by systematic errors in the stellar parameters, the inadequacy of classical abundance analysis that use static one-dimensional model atmospheres and LTE line formation calculations, or a combination of them.

For most of our sample stars, systematic offsets in the $\teff$ and $\logg$ scales would make the \ion{Fe}{i} and \ion{Fe}{ii} abundances agree. For a solar-like star, for example, an increase of 100 K in $\teff$ leads to an increase of 0.07 dex in the \ion{Fe}{i} abundance and a decrease of 0.03 dex in the \ion{Fe}{ii} abundance. If the surface gravity is then increased by 0.1~dex the \ion{Fe}{i} abundance decreases by 0.01~dex while the \ion{Fe}{ii} abundance increases by 0.04 dex. Thus, after an increase of 100 K in $\teff$ and 0.1 dex in $\logg$, the \ion{Fe}{i} abundance increases by 0.06 dex while leaving the \ion{Fe}{ii} abundance nearly unchanged. Excluding the cool metal-rich dwarfs, this would `solve' the ionization balance problem shown in Fig.~\ref{f:fe1vsfe2}.

Interestingly, a systematic difference of 100 K between the IRFM and the excitation balance temperature scales has been reported in the literature (Sect.~\ref{s:teff}). Also, on average, the ionization equilibrium gravities seem to be larger than those obtained using \textit{Hipparcos} parallaxes by roughly 0.1~dex (Allende~Prieto et~al. 1999). Note that if the effective temperatures are 100 K hotter, the trigonometric $\logg$ values would increase by 0.07 dex (cf. eq. 2 in Allende~Prieto et~al. 1999) so there is no need to question the accuracy of the \textit{Hipparcos} parallaxes in any case.

Alternatively, if the stellar parameters we adopted are correct, we should question the simplifications made in the abundance analysis, in particular the use of static plane parallel model atmospheres and LTE line formation. A full discussion of the effects of surface inhomogeneities and non-LTE effects on the Fe line formation is beyond the scope of this paper (see, e.g., Th\'evenin \& Idiart 1999, Asplund et~al. 2000, Shchukina \& Trujillo Bueno 2001, Korn et~al. 2003, Shchukina et~al. 2005). Nevertheless, evidence suggests that non-LTE effects may be responsible for the \ion{Fe}{i} vs. \ion{Fe}{ii} discrepancy.

Ionization by UV photons decreases the number density of neutral Fe atoms and increases that of \ion{Fe}{ii} ions. In warm stars ($\teff>5500$~K), most of the Fe is in the ionized form so the main effect on the abundances is the decrease of the \ion{Fe}{i} density. Thus, in a statistical equilibrium calculation where overionization by UV photons is the main non-LTE effect, the \ion{Fe}{i} lines are weaker than those predicted in LTE. In that case, to match the observations, the \ion{Fe}{i} abundance has to be increased compared to the LTE case, thus reducing the difference with the \ion{Fe}{ii} abundances, which do not change significantly due to this effect. Note, however, that this is expected to be less important in metal-rich cool stars, which contradicts the fact that the \ion{Fe}{i} vs. \ion{Fe}{ii} discrepancy worsens for our K-dwarfs of solar and super-solar metallicities (Fig.~\ref{f:fe1vsfe2}). Other model uncertainties, for example granulation or activity-related effects, may account for these large differences. Also, due to line blanketing, the overionization effect described above may result in a stellar atmosphere with a temperature structure different from that obtained in LTE, which may be also responsible for the differences seen at high $\feh$, as suggested by Feltzing \& Gustafsson (1998).

\subsubsection{Adopted iron abundances} \label{s:adoptedfe}

Due to the ionization balance problem described above, instead of averaging the abundances from \ion{Fe}{i} and \ion{Fe}{ii} lines, we would prefer to use the mean abundances of iron from the \ion{Fe}{ii} lines only as our $\feh$ scale due to their relatively low sensitivity to the atmospheric parameters adopted for most of our sample stars, i.e. those hotter than 5400~K, and small predicted non-LTE effects (e.g., Th\'evenin \& Idiart 1999, Gehren et~al. 2001).

Unfortunately, errors in the atomic data and the lower number of \ion{Fe}{ii} lines employed in this work compared to the number of \ion{Fe}{i} lines make the \ion{Fe}{i} abundances (most likely incorrect) \textit{internally} more accurate. In fact, the mean line-to-line scatter of the \ion{Fe}{i} abundances is 0.06~dex while that of the \ion{Fe}{ii} abundances is 0.10~dex. Thus, instead of discarding the \ion{Fe}{i} abundances, we prefer to embrace the scale of the \ion{Fe}{ii} abundances by increasing the \ion{Fe}{i} abundances by 0.06~dex. The adopted $\feh$ value is then a weighted mean of the modified \ion{Fe}{i} abundance and the \ion{Fe}{ii} abundance, where the weights are the individual values of the line-to-line scatter. Since the 0.06 dex difference is not constant with stellar parameters, we applied this correction only to stars with $\teff>5400$~K, $\logg>3.7$ (extrapolation to higher metallicities was allowed so the reader should be careful when using our $\feh$ values for stars with $\feh>0.1$).

\begin{figure}
 \includegraphics[width=8.5cm]{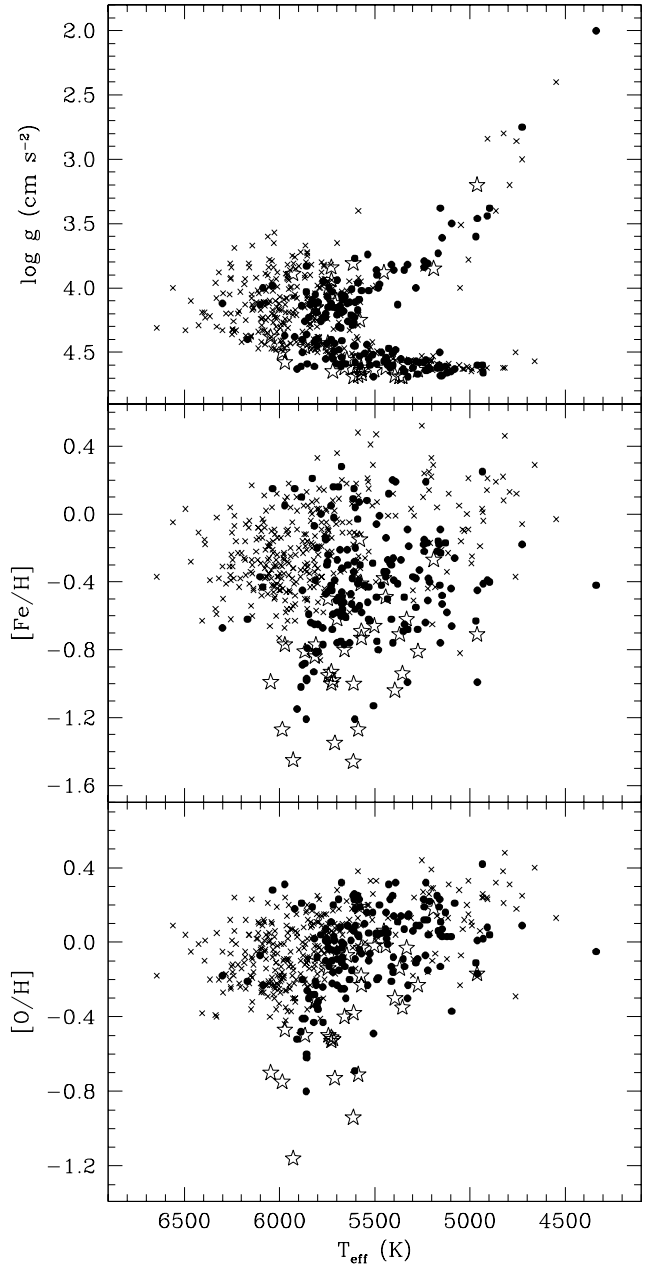}
 \caption{Upper panel: $\teff$ vs. $\logg$ for our sample stars. Middle panel: $\teff$ vs. $\feh$, as derived from the \ion{Fe}{ii} lines. Lower panel: $\teff$ vs. $\oxh$, where $\oxh$ has been non-LTE corrected. Crosses: $P_1>0.5$ (thin-disk stars), circles: $P_2>0.5$ (thick-disk stars), stars: $P_3>0.5$ (halo stars).}
 \label{f:hr}
\end{figure}

\bigskip

The adopted atmospheric parameters and derived iron and oxygen abundances are given in Table~6, available at the CDS. Figure \ref{f:hr} illustrates the overall properties of our sample in the HR diagram, as well as abundance vs. $\teff$ diagrams.

\section{Non-LTE corrections for the \ion{O}{i} triplet} \label{s:nlte}

\begin{figure*}
 \centering
 \includegraphics[width=13cm]{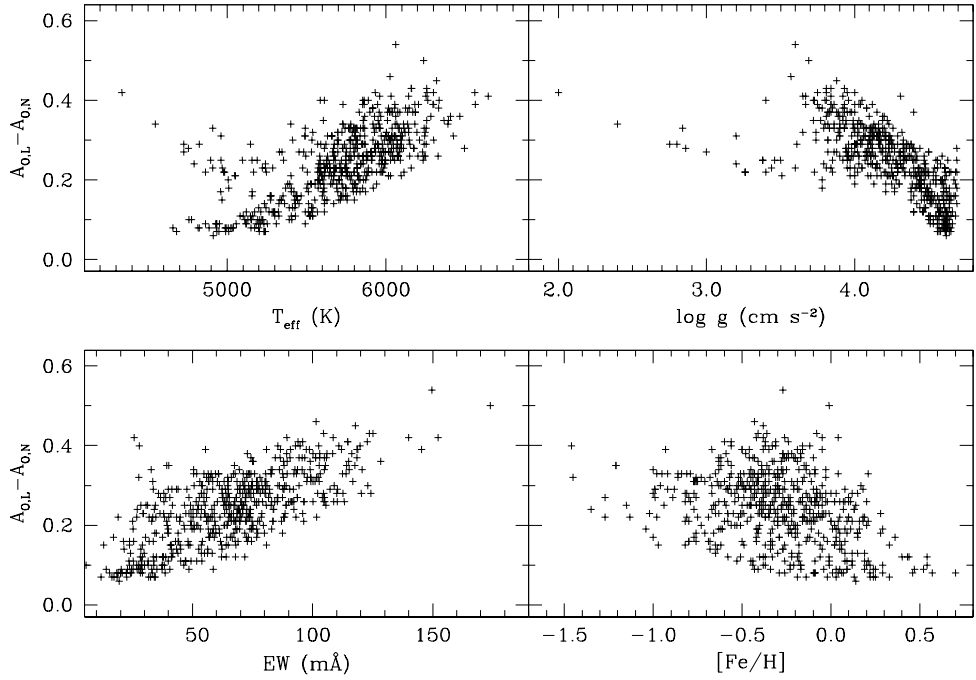}
 \caption{Difference between the oxygen abundance derived using LTE ($A_\mathrm{O,L}$) and non-LTE ($A_\mathrm{O,N}$) level populations, i.e. non-LTE corrections, for the middle line of the 777 nm \ion{O}{i} triplet as a function of atmospheric parameters and equivalent width.}
 \label{f:nltecorr}
\end{figure*}

The \ion{O}{i} triplet lines are known to be formed under conditions far from the LTE approximation, both in giant (Eriksson \& Toft 1979) and main sequence stars (Kiselman 1993, 2001). Our non-LTE corrections were calculated using fixed temperature and electron density structures from Kurucz LTE model atmospheres. Only the level populations of the hydrogen and oxygen atoms were allowed to depart from LTE by solving the rate equations and consistently recalculating the radiation field with TLUSTY.

We used an oxygen model atom with 54 levels plus the continuum and 242 transitions. Details on the atomic data used in the non-LTE calculations are given by Allende~Prieto et~al. (2003a). In particular, note that the model atom does not consider the fine structure of the triplet. Allende~Prieto et~al. (2003b) used this model atom to show that the lower level of the triplet is slightly overpopulated with respect to the LTE case while the upper level is noticeably underpopulated in the solar case. They also showed that the levels corresponding to the [\ion{O}{i}] 630 nm line do not depart substantially from LTE. These results are fairly consistent with those obtained by other authors (e.g., Kiselman 1993, Takeda 1994, Gratton et~al. 1999).

Once our calculations were complete, we discovered an error in the collisional ionization rates.\footnote{A corrected model atom has been made available at {\tt http://hebe.as.utexas.edu/at}} Nevertheless, the difference between the predicted equivalent widths ($EW$) of the triplet lines computed using the old and new model atoms is small. In general, the predicted $EW$s decrease by about 3\% and only a small dependence with temperature and metallicity is observed. The predicted $EW$s decrease by 3.2\% for a cool ($\teff=5000$~K) star and by 2.4\% for a hot star ($\teff=6400$~K) while they decrease by 2.8\% in the solar case and by 3.8\% for a $\feh=-0.8$ star of solar temperature. In all cases, the effect on the oxygen abundance ($A_\mathrm{O}$) derived by comparing the predicted $EW$s with the observed ones was an increase of 0.02~dex (rounded to the second decimal point). Therefore, the $\oxh$ ratios are unaffected given that the same abundance shift needs to be applied to our reference star (the Sun). 

The non-LTE corrections we calculated are shown in Fig.~\ref{f:nltecorr}. They are larger for warmer and low surface gravity stars in which the radiation field is stronger and the densities lower thus increasing the importance of radiative transitions and decreasing that of collisional excitation. More elegant descriptions of the non-LTE effects involved in the formation of the triplet can be found in Eriksson \& Toft (1979) and Kiselman (1993).

An important but still uncertain ingredient for the non-LTE corrections are inelastic collisions with neutral H atoms. If included, the non-LTE corrections will be reduced. Allende~Prieto et~al. (2004b) have shown that H collisions, although having a marginal and usually neglected role at solar metallicity (e.g., Nissen et~al. 2002), help to reproduce better the center-to-limb variation of the triplet line profiles when included using Drawin's formula (Drawin 1968) multiplied by an empirical factor ($S_H$), as suggested by Steenbock \& Holweger (1984). Note, however, that this effect becomes more important at low metallicities (Ram\'{\i}rez et~al. 2006). 

The key effect of introducing H collisions is the decrease of thermalization with depth due to the decrease of neutral H density, which leads to smaller non-LTE corrections for the strongest line of the triplet compared to those for the weakest line (see Allende~Prieto et~al. 2004b and Ram\'{\i}rez et~al. 2006).

\begin{figure*}
 \centering 
 \includegraphics[width=18.0cm]{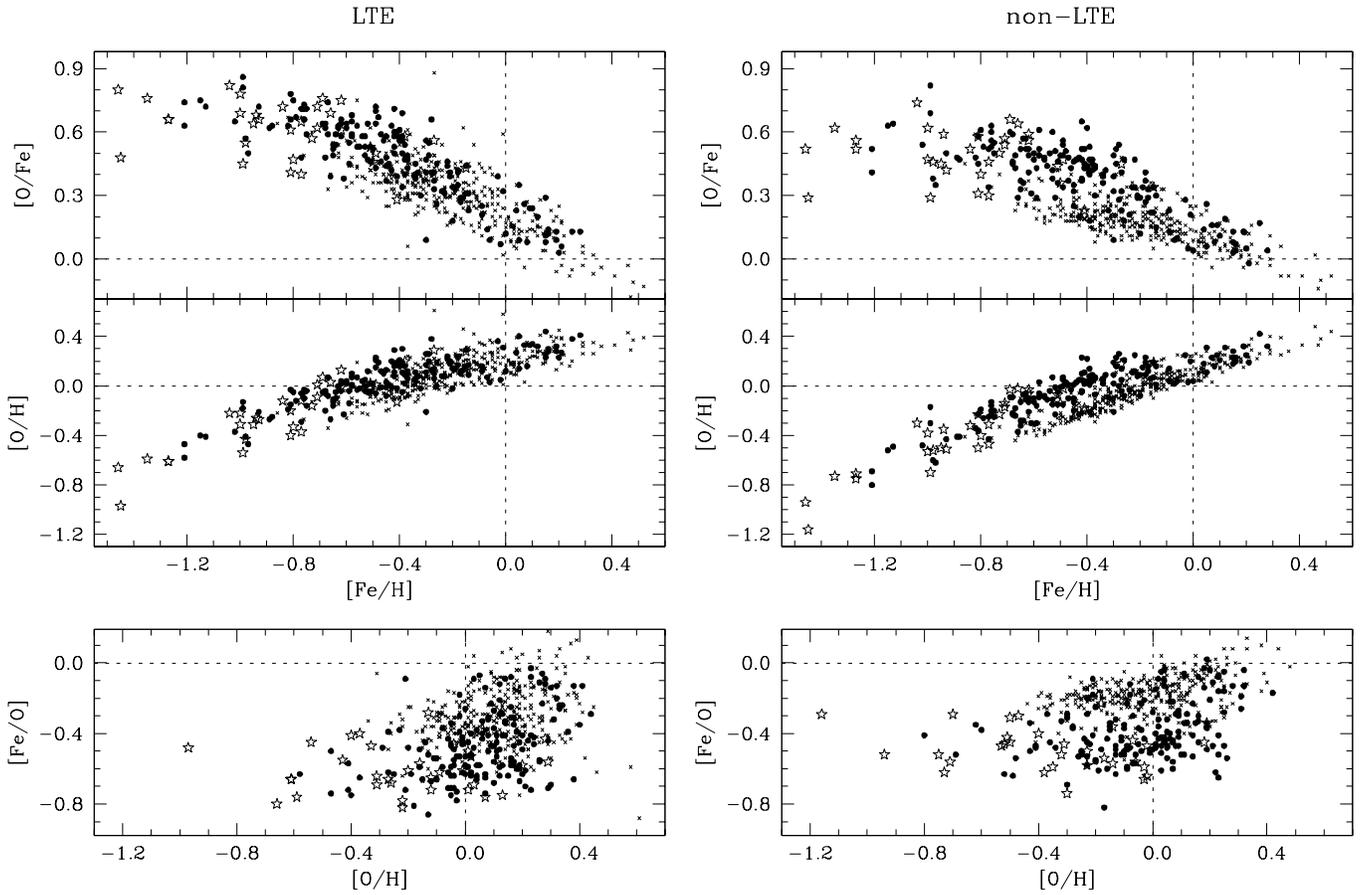}
 \caption{Oxygen vs. iron abundance trends for all our sample stars before (left side panels) and after (right side panels) applying our non-LTE corrections. Crosses: $P_1>0.5$ (thin-disk stars), circles: $P_2>0.5$ (thick-disk stars), stars: $P_3>0.5$ (halo stars). The $\feh$ values are from the \ion{Fe}{ii} lines only.}
 \label{f:ofe1}
\end{figure*}

Our non-LTE calculations do not include the effect of H collisions. In this way, by analyzing our results we find that in order to obtain consistent non-LTE abundances from the three lines of the triplet, the non-LTE corrections of the 7771.9~\AA\ and 7774.2~\AA\ lines have to be decreased by 0.036~dex and 0.018~dex, respectively, to match the abundance from the 7775.4~\AA\ line (nearly independent of stellar parameters). These mean empirical corrections were applied to force an agreement between the abundances derived from each line of the triplet, which in turn reduced the scatter in the oxygen abundance patterns but may be still giving systematically low abundances, with, perhaps, a trend with temperature and metallicity. Note that, for example, Allende~Prieto et~al. (2004b) find an increase in the solar O abundance of 0.06 dex when going from a non-LTE calculation without H collisions ($S_H=0$) to one with them ($S_H=1$) while Ram\'{\i}rez et~al. (2006) find an increase of 0.25~dex for a star with $\feh=-1.7$ when going from $S_H=0$ to $S_H=10$.\footnote{In each case, the $S_H\ne0$ values were chosen to reproduce accurately the triplet line profiles.}

For the Sun, we obtained a mean non-LTE correction of 0.13 dex, which leads to a non-LTE oxygen abundance of $A_\mathrm{O}=8.72\pm0.02$, in agreement with other recent solar oxygen abundance determinations, as mentioned in Sect.~\ref{s:ldandsolar}. 

As in the case of the iron abundances, for the sample stars, oxygen abundances were calculated differentially on a line to line basis with respect to the Sun, and are given as [O/H] in Table~6, available at the CDS.

\subsection{A grid of non-LTE corrections} \label{s:nltegrid}

The non-LTE abundances given in Table~6 were computed `directly' from curves of growth corresponding to the stellar parameters of each particular star. For practical purposes, we also computed a grid of non-LTE corrections in the $\teff,\logg,\feh,A_\mathrm{O,L}$ space, which allows to obtain non-LTE corrections, given the atmospheric parameters and LTE oxygen abundances from each line of the triplet, for any star with parameters within those covered by our sample. The grid and an IDL routine to interpolate within this grid are available from the authors upon request. The non-LTE abundances obtained directly and those inferred using the grid agree at the 0.02~dex level without any trends with stellar parameters.

\section{Oxygen Abundance Patterns} \label{s:ofetrends}

\begin{figure*}
 \centering
 \includegraphics[width=12cm]{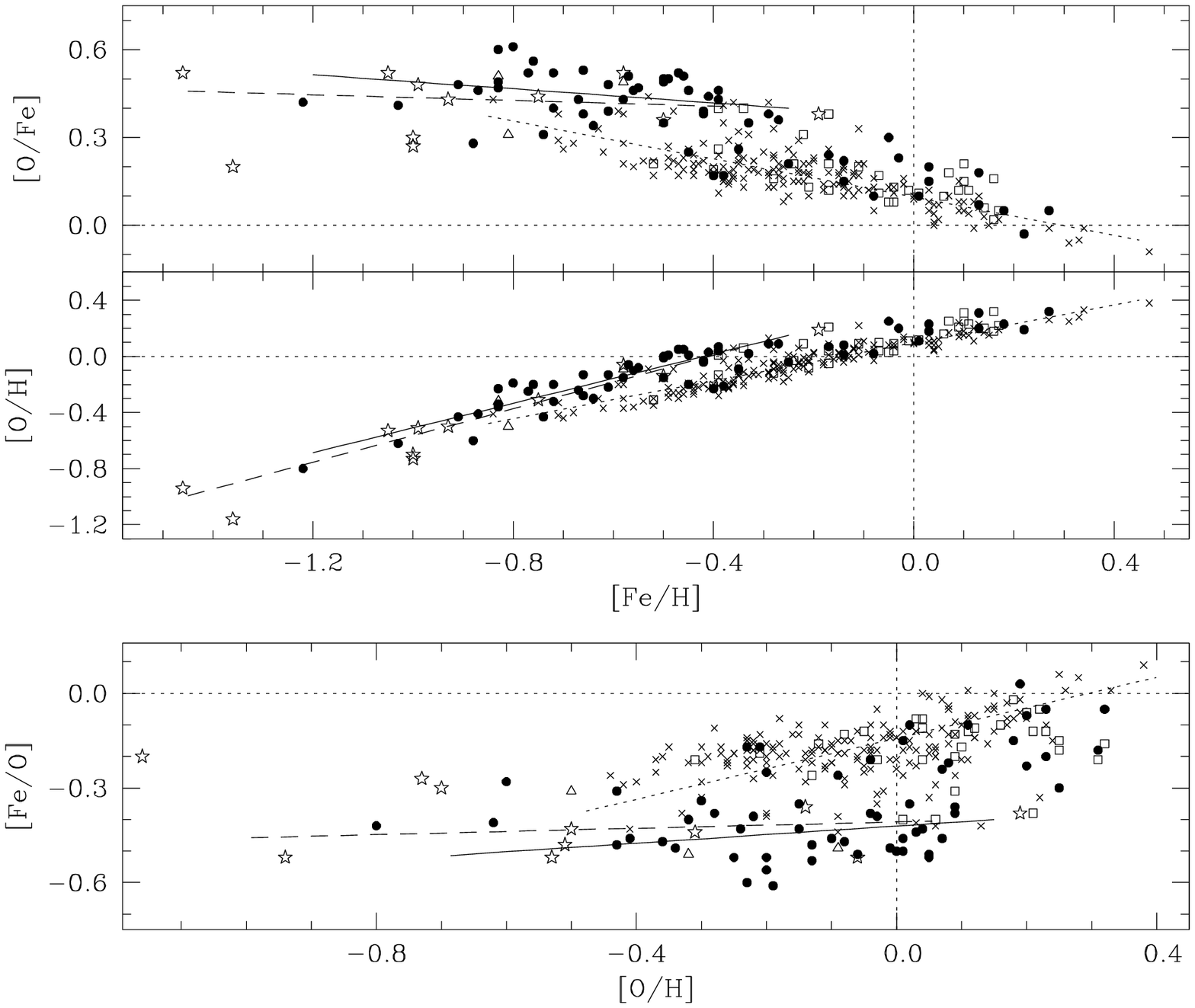}
 \caption{As in the right side panels of Fig.~\ref{f:ofe1} but restricted to stars with $\teff>5400$~K, $\logg>3.8$, and membership probabilities larger than 70\%. The $\feh$ values here are a mean of the \ion{Fe}{ii} and modified \ion{Fe}{i} abundances (Sect.~\ref{s:adoptedfe}). Squares and triangles represent stars whose kinematic probabilities of being either thin-disk, thick-disk, or halo stars are lower than 70\%. Triangles are stars that, kinematically, may belong to either the thick disk or halo, while squares represent stars that may belong to the thin disk or the thick disk. The dotted, solid, and dashed lines are linear fits to the thin-disk, thick-disk, and halo samples, respectively (see Table~\ref{t:trends}), when the kinematic probabilities are larger than 70\%. For the thick-disk fit, we excluded the stars that having thick-disk kinematics follow the thin-disk abundance pattern.}
 \label{f:ofe2}
\end{figure*}

Here we describe the oxygen vs. iron abundance trends we derived for halo, thin- and thick-disk stars using the preceding results. Interpretations of the results are given in Sect.~\ref{s:discussion}.

Fig. \ref{f:ofe1} shows the trends we obtained before and after applying the non-LTE corrections to all our sample stars. It is clear that the non-LTE corrections reduce the scatter and make the thin/thick disk differences obvious compared to the LTE case. In the LTE case, the [O/Fe] vs. [Fe/H] relation shows a steeper slope compared to the non-LTE case. Also, in the LTE case, no clear differences are seen between thin- and thick-disk stars, except in the [Fe/O] vs. [O/H] plane, where an apparent continuous increase in the [Fe/O] ratios from thick- to thin-disk stars is seen at all [O/H]. The fact that the thin/thick disk separation becomes obvious after applying the non-LTE corrections suggests that our non-LTE calculations are reliable.

To reduce the impact of errors in the atmospheric parameters and model uncertainties, we also show the oxygen abundance trends for stars with $\teff>5400$~K and $\logg>3.8$ (i.e. dwarfs only) in Fig.~\ref{f:ofe2}. In this way, besides restricting the sample to stars with similar parameters, we also avoid the coolest dwarfs that suffer more strongly from the ionization balance problem discussed in Sect.~\ref{s:noionization}. In fact, the [Fe/H] values in Fig.~\ref{f:ofe2} correspond to those derived in Sect.~\ref{s:adoptedfe}, where the \ion{Fe}{i} abundances are empirically put into the more reliable \ion{Fe}{ii} abundance scale. Furthermore, in Fig.~\ref{f:ofe2} we discriminate between stars with high and intermediate probabilities of being thin-disk, thick-disk, or halo stars by adopting the strong constrain of $P_i>0.7$, which rules out any possibility for the errors in the $U,V,W$ velocities of a given star to allow $P_i$ to be lower than 0.5 (see Sect.~\ref{s:kinematic-criterion}).

\subsection{Halo, thin-disk, and thick-disk stars} \label{s:thin-ck}

We computed linear fits to the data shown in Fig.~\ref{f:ofe2}, separating the stars into halo, thin-disk, and thick-disk members using the strong constraint of 70\% in the kinematic membership probabilities. Also, the thick-disk fit does not include the kinematically-selected thick-disk members that follow the thin-disk abundance pattern. These are discussed separately in Sect.~\ref{s:knee}. The results of the fits are given in Table~\ref{t:trends}. For these calculations, the random errors in $\feh$ and $\ofe$ for each star were taken into account.

Although there are a few very metal-poor halo stars within 150 pc from the Sun, we restricted our sample of halo stars to $\feh>-1.5$ because our procedure, specifically the automatic measurement of equivalent widths (Sect.~\ref{s:ew}), works best at these metallicities. Therefore, our halo star sample is not complete, and any attempt to trace a halo abundance trend using only the metal-rich end is dubious. Nevertheless, there is an overlap in the $\feh$ distributions of halo and thick-disk stars from about $-1.2$ to $-0.4$ which can be analyzed using the results from this study. In this range the mean $\ofe$ ratios of thick-disk and halo stars are similar at about 0.45. By inspecting the linear fits in Table~\ref{t:trends}, however, we notice that only the thick disk trend shows a significant negative slope. Halo stars seem to have, statistically, constant $\ofe$ in this metallicity range while the $\ofe$ ratio of thick disk stars decreases by almost 0.1~dex. Note also that the star-to-star scatter for the halo star sample is slightly larger than that of the thick disk.

A quick inspection of Figs.~\ref{f:ofe1} and \ref{f:ofe2} reveals that the majority of thick-disk stars have larger [O/Fe] and [O/H] ratios compared to thin-disk stars for $\feh<-0.2$. Ignoring the thick-disk stars that appear to follow the thin-disk abundance trend, we conclude from Figs.~\ref{f:ofe1} and \ref{f:ofe2} that there are two different and mostly separated abundance patterns for thin- and thick-disk stars. Compared to thin-disk stars, thick-disk members have larger [O/Fe] and [O/H], and smaller [Fe/O] in the ranges $-0.9<\feh<-0.2$, $-0.5<\oxh<0.3$, respectively. The differences are roughly constant at about 0.2 dex in [O/H] and 0.3 dex in [Fe/O], but they depend on [Fe/H] in the case of [O/Fe] because the thin disk [O/Fe] ratio has a steeper slope (Table~\ref{t:trends}). Note also that while the thick disk trend is well fitted by a line, the thin disk trend seems to require a higher order polynomial to satisfactorily fit the data.

\begin{table}
\centering
\begin{tabular}{lccrc} \hline\hline
           &             $a$ &              $b$ & $n$ & $\sigma$ \\ \hline
Thin disk  & $0.096\pm0.004$ & $-0.327\pm0.016$ & 238 &   0.056 \\
Thick disk & $0.370\pm0.027$ & $-0.121\pm0.043$ &  43 &   0.065 \\
Halo       & $0.388\pm0.049$ & $-0.048\pm0.071$ &  12 &   0.072 \\ \hline
\end{tabular}
\caption{Linear fits to the data shown in Fig.~\ref{f:ofe2}. Given here are the coefficients of the fit: $\ofe=a+b\feh$, the number of stars used ($n$) and the random scatter around the mean fit ($\sigma$). The linear fits and the $\sigma$ computation take into account the individual errors in $\feh$ and $\ofe$.}
\label{t:trends}
\end{table}

We estimate the random error due to mean errors of 50~K in $\teff$ and $0.05$~dex in $\logg$ to be about 0.03~dex for $\feh$, the line-to-line scatter is 0.06 for $\feh$ and 0.03 for $\oxh$ after correcting for non-LTE, so the random error in $\ofe$ should be about 0.07 dex. We expect this random error to be slightly larger for the halo sample due to uncertainties in reddening, parallax, and lower $S/N$ spectra. It is also reasonable to conclude that the scatter is smaller for the thin-disk sample, for which better spectra are available, as well as accurate stellar parameters. Thus, we find that there is little or no room for cosmic scatter in the thin-disk and  the thick-disk abundances, in agreement with the results of R03 and R06. Although we can be tempted to draw the same conclusion for the halo sample, we should be cautious and warn the reader that, before that, more metal-poor stars need to be added to the sample. Note that, for example, Mel\'endez et~al. (2006) find that the observed scatter in the $\ofe$ ratios of halo stars in the $-3<\feh<-1$ range allows for a small intrinsic scatter.

\subsection{An intermediate population?} \label{s:interm}

\begin{figure}
 \centering
 \includegraphics[width=6.5cm]{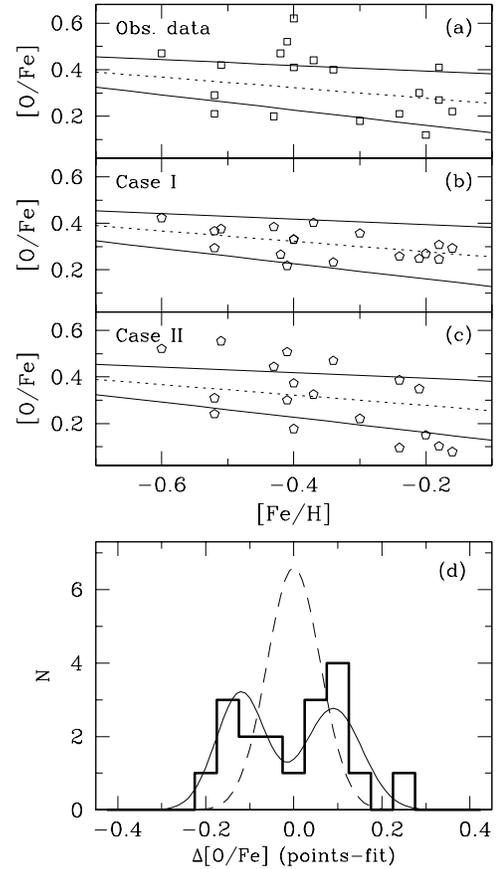}
 \caption{(a) Stars in our sample with low kinematic probability ($<70\%$) of being either thin- or thick-disk stars in the [O/Fe] vs. [Fe/H] plane. The solid lines represent the mean trends of thin- and thick-disk stars with high kinematic probabilities in our sample, as given in Table~\ref{t:trends}. The dotted line represents the intermediate abundance trend of a ficticious population with chemical properties between those of the thin and thick disk. (b) Case I: simulated stars (18, the same number of stars as in (a)) assuming that they all belong to the intermediate population. (c) Case II: simulated stars assuming that half (9) belong to the thin disk and half (9) to the thick disk. (d) Scatter distribution of the observed (panel (a), thick solid line histogram) and simulated [Case I (panel (b), dashed line), Case II (panel (c), thin solid line)] data. Gaussian fits are used for the simulated data.}
 \label{f:interm}
\end{figure}

Our sample is biased towards stars with high kinematic probabilities of being either thin- or thick-disk stars because it is dominated by nearby stars (hence the large number of thin-disk members) and a sample selection was specifically made to collect a significant number of stars with kinematics similar to those of the thick disk. Thus, only a few stars (18) in our sample have low kinematic probabilities of being either thin- or thick-disk stars and only these should be looked at to check if an intermediate population with properties between those of the two disks exists. These 18 stars are shown in Fig.~\ref{f:interm}a.

A quick inspection of Fig.~\ref{f:interm}a (see also Fig.~\ref{f:ofe2}) already shows that stars with a kinematics intermediate between that of the thin and thick disk seem to clearly follow either the thin disk or the thick disk abundance pattern and not an intermediate one. To show this quantitatively, we simulated the scatter around the mean trends for the stars in our sample that have kinematic probabilities of being thin- or thick-disk stars below 70\% assuming that: (I) they belong to a population with an abundance pattern between those of the thin and thick disk (Fig.~\ref{f:interm}b), and (II) they belong to either the thin disk or the thick disk (Fig.~\ref{f:interm}c). The mean thin/thick disk trends and scatter adopted here are those given in Table~\ref{t:trends}. For the intermediate population we assumed an scatter of 0.06~dex, which is the average of the scatter observed in thin- and thick-disk stars.

Clearly, case (I) provides a poor representation of the data. Even if we assume that the intermediate population has a larger scatter around the mean abundance trend, it would be hard to explain the dip in the scatter distribution shown in Fig.~\ref{f:interm}d. On the other hand, case (II) satisfactorily reproduces the dip and the peaks in Fig.~\ref{f:interm}d even though the statistics is poor (only 18 points). A Kolmogorov-Smirnov test (see, e.g., Press et~al. 1992) shows that the probability that the data shown in Fig.~\ref{f:interm}a belong to the double peaked thin/thick disk distribution (the thin solid line in Fig.~\ref{f:interm}d) is 96\% while the significance level for the intermediate population case (the dashed line in Fig.~\ref{f:interm}d) is less than 1\%. Only if we assume that the scatter around the mean trend of the intermediate population is 0.14~dex, this number increases to a maximum of 72\%. In such case, we would need a cosmic scatter of about 0.12~dex in the intermediate population. Given that no cosmic scatter is seen in the thin or thick disks (Sect.~\ref{s:thin-ck}), this wide distribution seems unrealistic. This analysis discards the existence of a population of stars with abundances intermediate between those of the thin and thick disk, at least in the solar neighborhood.

\subsection{The `knee' and Reddy et~al. TKTA stars} \label{s:knee}

In a series of papers, T. Bensby and collaborators have claimed that the abundance trends of the thick disk show the contribution of Type Ia supernovae (Feltzing et~al. 2003; Bensby et~al. 2003, 2004, 2005). In the case of oxygen, Bensby et~al. (2004, hereafter B04) found that from an almost constant $\ofe=0.35$ at low metallicities, thick disk stars with $\feh>-0.3$ show $\ofe$ ratios that decrease rapidly towards $\ofe=0$ at $\feh=0$, i.e. they find a `knee' in the $\ofe$ vs. $\feh$ relation of thick-disk stars.

\begin{figure*}
 \centering
 \includegraphics[width=13cm]{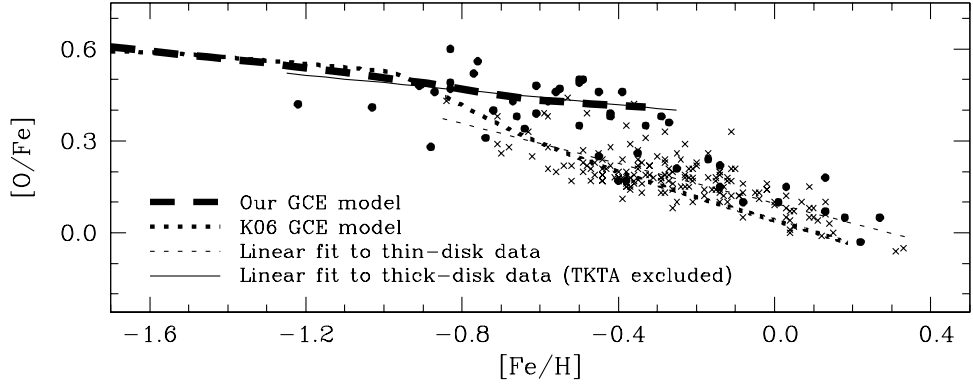}
 \caption{As in Fig.~\ref{f:ofe2} but restricted to thin- and thick-disk stars with $\teff>5600$~K and membership probabilities larger than 70\%. The dotted and solid lines correspond to the mean trends given in Table~\ref{t:trends} using a larger sample, the thick dashed line shows the predictions of our simple GCE model for the thick disk (the details of the model are given in the Appendix~\ref{s:gce} and the discussion in Sect.~\ref{s:discussion}), and the thick dotted line is the model by Nomoto et~al. (2006) shifted upwards by 0.1~dex to match our results at low metallicity.}
 \label{f:ofe3}
\end{figure*}

In Fig.~\ref{f:ofe2} there is a significant number of stars with high kinematic probability of being thick-disk stars having $\feh>-0.4$. Some of these were recently observed by us and are thus not included in the Reddy et~al. papers. We targeted this sample aiming at a better definition of the knee and/or determine where the thick disk [Fe/H] distribution ends. However, we found that almost all of these stars follow the thin disk abundance pattern. In fact, in Fig.~\ref{f:ofe1} we also see some thick-disk stars at even lower metallicities that follow the thin disk abundance pattern. This suggests that the kinematics alone does not always correctly separates thin- from thick-disk members. The possibility of errors in the abundance determination for these outliers is very unlikely. Therefore, following R06 designation (see their Sect.~6.1), these are TKTA stars (stars with Thick-disk Kinematics but Thin-disk Abundances).

The TKTA stars are probably either old thin-disk stars that have been heated by secular processes (more than the average) or members of a moving group that can be ultimately associated with the thin disk. As shown by R06, the TKTA stars are, on average, older than a typical thin-disk member but apparently younger than a typical thick-disk star and their distribution in the $U,V,W$ galactic velocity space shows a preference towards positive $U$ velocities even though thick-disk members cover both positive and negative $U$ velocities, more or less uniformly, at least in the $\pm150$ km~s$^{-1}$ range. All the stars we classify as TKTA are older than 4~Gyr (Fig.~\ref{f:ages}) but their distribution in $U,V,W$ is more random compared to the R06 case which implies that only some TKTA stars (most of them in the R06 sample) may be members of a moving group. Note also that the Galactic bar may have affected the orbits of some low eccentricity stars in such a way that they now resemble thick-disk members, as suggested by Soubiran \& Girard (2005). This may also explain the existence of stars with low kinematic probabilities of being thin- or thick-disk members, which, instead of being members of an intermediate population, show clearly thin-disk abundances (Sect.~\ref{s:interm}).

Two thick-disk stars in Fig.~\ref{f:ofe2} with $\feh$ slightly below solar do show clearly higher [O/Fe] ratios compared to the mean thin disk (HIP\,99224 and HIP\,90542). They both have $\teff\simeq5500$~K, closer to our lower limit of 5400~K, a temperature at which the triplet lines begin to give systematically large abundances for young cool dwarf stars (Morel \& Micela 2004, Schuler et~al. 2006), most likely due to the use of classical model atmospheres that do not include activity related and/or granulation effects. Note also that this is roughly the temperature at which the ionization balance problems described in Sect.~\ref{s:noionization} begin to become important. If we restrict the sample to stars with $\teff>5600$~K, the `oxygen-enhanced' thick disk ends at $\feh=-0.3$, again without obvious evidence for a knee (see Fig.~\ref{f:ofe3}).  Therefore, our sample, expanded to better define the metal-rich end of the thick disk, shows no signature of a knee in the thick disk [O/Fe] vs. [Fe/H] relation. Given that others have proposed that metal-rich thick-disk stars exist and define a knee (in particular Bensby et~al. 2004), our solid conclusion from this is that not all metal-rich stars with thick disk kinematics show that behavior. In fact, the vast majority of them do not follow the knee, if there is any at all.

\subsection{Re-analysis of the Bensby et~al. data}

\begin{figure}
 \centering
 \includegraphics[width=7cm]{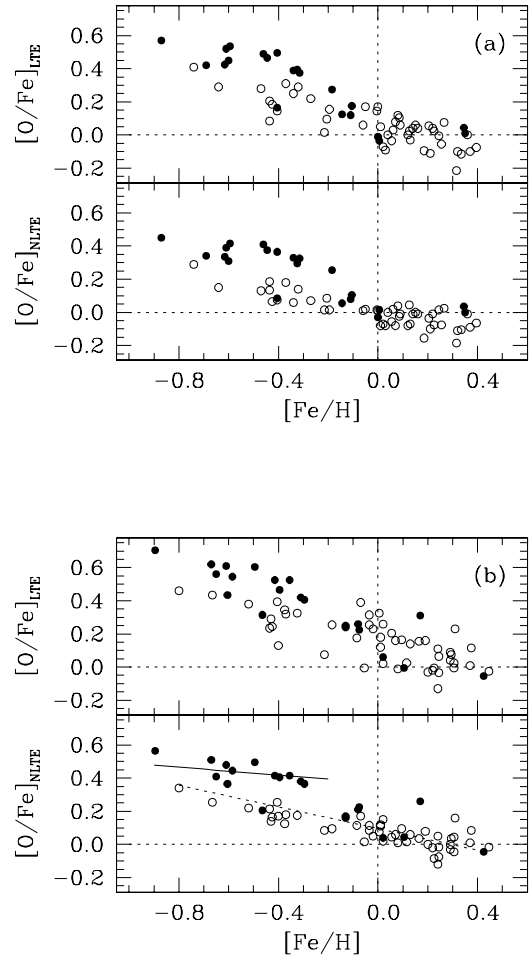}
 \caption{(a) LTE and non-LTE Abundance trends for the thin- (open circles) and thick-disk (filled circles) stars in Bensby et~al. (2004) using their atmospheric parameters but our atomic line data and non-LTE corrections. (b). As in (a) but using our $\teff$ and $\logg$ scales. The dotted and solid lines are the mean trends found with our sample (Table~\ref{t:trends}).}
 \label{f:bensby}
\end{figure}

The equivalent widths for the lines used in B04 were kindly supplied by T. Bensby. Using these data and the atmospheric parameters from their paper, we obtained LTE iron and oxygen abundances, and corrected for non-LTE effects using our grid (Sect.~\ref{s:nltegrid}). The results are shown in Fig.~\ref{f:bensby}a.

Our LTE oxygen abundance trends, when using the B04 parameters, are essentially the same as those obtained by B04 (see their Fig. 9c), as expected. The small differences are only due to the use of slightly different $\log gf$ values. When corrected using our non-LTE calculations, no clear evidence of a knee is observed, in contrast with the B04 non-LTE result, which was obtained using non-LTE corrections from Gratton et~al. (1999). Comparing their Fig.~9d with our Fig~\ref{f:bensby}a we notice that it is essentially one star (HIP 109450: $\feh\sim-0.15$ in B04, $\feh\sim-0.10$ in our work) that `defines' the knee in the B04 work (using the triplet, their knee obtained using the results from the forbidden 630~nm [\ion{O}{i}] line is defined by more stars). This star has a near solar temperature but it is slightly more evolved ($\logg\sim4.2$). Thus, its non-LTE correction should be larger than that of the Sun by about 0.1~dex (Fig.~\ref{f:nltecorr}) leading to a non-LTE [O/Fe] ratio smaller by roughly the same amount when compared to the LTE case. However, a visual inspection of Fig.~9d of B04 suggests that their LTE and non-LTE [O/Fe] ratios for this star are very similar. Therefore, the non-LTE corrections adopted by B04 may not be correctly taking into account the effects of the surface gravity. Our Fig.~\ref{f:bensby}a shows that the [O/Fe] ratio of this star has indeed decreased by 0.1~dex when going from LTE to non-LTE. As a result, in our Fig~\ref{f:bensby}a, HIP~109450 is clearly following the thin disk abundance pattern, contrary to the B04 case, in which the star is in between the thin- and thick-disk trends, thus creating an apparent knee.

In Fig~\ref{f:bensby}b we show the abundance trends obtained with the B04 data but using effective temperatures, metallicities, and surface gravities computed as in our work. The systematic differences in the abundances, compared to B04, are mainly due to the use of different temperature scales. When non-LTE corrected, the abundance trends derived from the B04 data are in excellent agreement with those found in our large sample (Table~\ref{t:trends}), shown in Fig.~\ref{f:bensby}b with solid lines. This shows that the differences with the B04 work for the triplet lines, in particular the discrepancy regarding the existence of the knee, are not a consequence of sample selection or data quality.

Note that Bensby et~al. (2005) add more stars to the B04 sample, thus strengthening their O vs. Fe trends (including some `knee stars') obtained using the 630 nm [\ion{O}{i}] line. Unfortunately, the spectral region where the triplet is was not included in their latest observations (T. Bensby, private communication) and thus no further comparisons are possible.

\section{Stellar ages} \label{s:ages}

To complement the observational evidence, we computed the ages of our sample stars using our derived stellar parameters and the Bertelli et~al. (1994) isochrones, as explained in R03 and AP04a. These ages may be subject to large systematic uncertainties, as explained in the Appendix~\ref{s:ageerrors}, but in order to maintain consistency with the other stellar parameters derived from the same set of isochrones (namely the $\log g$ values), we did not correct for those.

\begin{figure}
 \includegraphics[width=8.5cm]{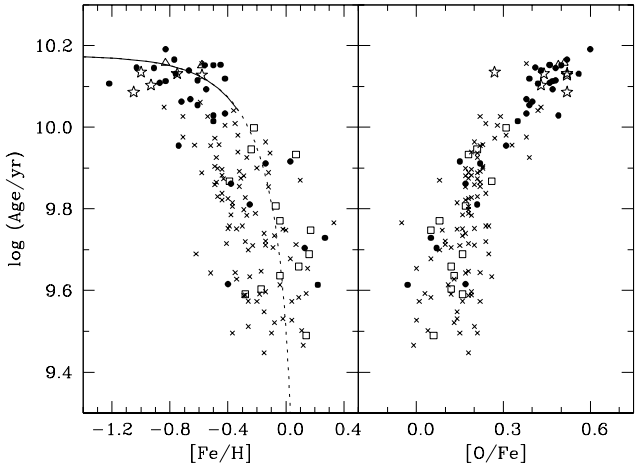}
 \caption{Age vs. abundance for the stars shown in Fig.~\ref{f:ofe2} for which reasonably accurate ages can be derived. The dashed line corresponds to the age-metallicity relation adopted in our simple GCE model (Appendix~\ref{s:gce}) with the solid line showing the first 4 Gyr of that relation.}
 \label{f:ages}
\end{figure}

In Fig.~\ref{f:ages} we plot the stellar ages vs. abundances for the stars in Fig.~\ref{f:ofe2}. Only the stars for which precise ages could be determined ($\sigma(\log \mathrm{(Age/yr)})<0.3$~dex, where $\sigma$ is the internal uncertainty) are shown.\footnote{The lower and upper limits in the age of a given star were calculated from its age probability distribution (APD). The lower (upper) limit corresponds to the value at which the integrated APD, starting from the peak, reaches 0.955/2, i.e., the equivalent of the area covered by a Gaussian APD from the peak to the lower (upper) 2-$\sigma$ limit. This procedure results in more realistic error bars given the asymmetry of the APDs.} It is clear that our thick-disk stars with $\ofe>0.25$ are older than our thin-disk stars and that the former seem to cover a short age range (10--14~Gyr) while thin-disk stars as young as 2.5~Gyr or less and as old as 10 Gyr are found. These ages are in reasonable agreement with those derived by Fuhrmann (1998), who suggests that all thin disk stars are younger than 9 Gyr while all thick disk stars are older than 10~Gyr, and Gratton et~al. (2000, their Figs. 5d and 6d). An interesting conclusion may be drawn from these results: the thick disk was formed before the thin disk, or at least the bulk of it. Whether this implies a hiatus of star formation, and therefore an age gap in Fig.~\ref{f:ages}, between the formation of the thick and thin disks (as suggested by Fuhrmann 1998 and Gratton et~al. 2000 based on the abundance information) depends on the accuracy of the ages. An age gap might be present, but the uncertainties in our derived ages make difficult to detect it. We remark that the majority of the young stars ($\mathrm{Age}<8$~Gyr) identified, kinematically, as thick-disk members in Fig.~\ref{f:ages} follow the thin-disk abundance pattern. 

\section{Discussion} \label{s:discussion}

Here we discuss the implications of our results for thin- and thick-disk stars. Our halo star sample is by far incomplete so we will not discuss the relevance of our results for halo studies.

Oxygen is believed to be produced mainly in massive stars and put into the interstellar medium (ISM) by Type~II supernovae (SN~II) explosions, while iron is produced in both SN~II and Type~Ia supernovae (SN Ia) events. The time scale for SN~II explosions to occur is about $3\times10^7$ yr, which roughly corresponds to the lifetime of a star with a mass of $8M_\odot$. For a SN~Ia event, however, one must wait for a low or intermediate mass star to evolve into a white dwarf and accrete material from a companion. Since lifetimes of these stars are substantially longer, the time scale for the first SN Ia events to take place is also longer, of the order of $\sim10^9$~yr, although it should be noted that the exact value for the time scale of SN~Ia events is very uncertain (see, e.g., Greggio 2005 for a recent discussion on this topic). In principle, stellar oxygen and iron abundances, as well as their ratios, contain information about the chemical enrichment of the ISM. This simple picture is complemented by additional ingredients such as the supernova yields that determine the net amount of the different elements ejected in the explosions, and the star formation rate and initial mass function (e.g., Wheeler et~al. 1989, McWilliam 1997).

Intermediate mass stars (those with masses between about 1 and 10 solar masses) also contribute a significant fraction of metals to the ISM (e.g., Iben \& Renzini 1983). During their evolution, in the so-called Asymptotic Giant Branch (AGB) phase, these stars undergo a series of thermal pulses that cause heavy element production (particularly C, N, and $s$-process elements) in the helium burning shells, which are later ejected with the envelope as winds and superwinds.

Note that the stellar yields may strongly depend on the initial metallicity and the minimum mass for black hole formation (e.g., Schild \& Maeder 1985, Maeder 1992). In the former case, metallicity affects stellar evolution through changes in the opacities, which affect both the internal structure and mass loss rates. Oxygen production seems to be favored at low metallicities (e.g., Kobayashi et~al. 2006). Regarding black hole formation, an uncertainty in the minimum mass required to form a black hole ($M_\textrm{\tiny{BH}}$) leads to an uncertainty in the net heavy element yield. If $M_\textrm{\tiny{BH}}$ is large, then more stars (those with lower masses) will effectively eject their outer layers rich in heavy elements. On the other hand, if $M_\textrm{\tiny{BH}}$ is small, more stars will become black holes and swallow part of the heavy elements during the collapse. Thus, the heavy element yields decrease with a decreasing $M_\textrm{\tiny{BH}}$. Note also that the \textit{net} yield depends on both the initial metallicity and minimum mass for black hole formation, given that some of the synthesized material is ejected by winds. While this is true for SN~II events that lead to black hole formation, in general the amount of metals returned to the ISM depends on the mass of the compact remnant, which may be also a neutron star.

The abundance trends we derive for the thin and thick disks (Figs.~\ref{f:ofe1}, \ref{f:ofe2}, and \ref{f:ofe3}) suggest that the gas from which thick-disk stars were formed was rich in material synthesized in massive stars, given the larger [O/H] and particularly [O/Fe], essentially at all [Fe/H]. The thick disk [O/Fe] ratio, which is, in a first approximation, roughly constant at about 0.5 dex, is consistent with the [O/Fe] ratio of about 3 times solar measured in the ejecta of the Type~II SN1987A (see, e.g., Arnett et~al. 1989). Thin-disk stars, on the other hand, formed from a gas in which a mixture of material from low mass and massive stars was present. The steep decline in the thin disk [O/Fe] ratio indicates that the chemical enrichment of the ISM was due to both SN~Ia and SN~II, as proposed originally by Tinsley (1980), and demonstrated here, in Fig.~\ref{f:ofe3}, by the satisfactory agreement between the data and the Galactic Chemical Evolution (GCE) model by Nomoto et~al. (2006).

Inspection of Figs.~\ref{f:ofe1}, \ref{f:ofe2}, and \ref{f:ofe3} reveals that the thick disk [O/Fe] ratios decrease from 0.5 to 0.4 with [Fe/H] from $\feh=-0.8$ to $-0.3$ and that no knee in the [O/Fe] vs. [Fe/H] trend is present at higher metallicities. The small slope can be explained by a metallicity dependent yield from SN II because oxygen production is favored at low $\feh$. Using the nucleosynthesis yields by Kobayashi et~al. (2006, hereafter K06) and a simple GCE model (see Appendix~\ref{s:gce}), we were able to satisfactorily reproduce the slope in the [O/Fe] vs. [Fe/H] relation of the thick disk (see Fig.~\ref{f:ofe3}). Only massive stars were allowed to enrich the ISM in our model so, given the good agreement with the data, we conclude that the SN~II to SN~Ia ratio during the formation of the thick disk was large enough to make the contribution of SN~Ia to the chemical enrichment of the ISM negligible. This can be achieved, for example, by a continuous intense star formation during the whole process of thick disk formation. Note that the lack of a knee does not exclude SN~Ia events and their contribution, albeit small, of metals to the ISM during the formation of the thick-disk.

The [O/Fe] vs. [Fe/H] trend in our simple GCE model is almost insensitive to the choice of star formation rate (e.g., continuous, truncated, or exponentially decaying).\footnote{Partially because we did not include SN~Ia.} The metallicity dependence of the oxygen yields by K06 lead to a good representation of our observed data. Note, however, that the predicted metallicity dependence of the yields for other elements might be incompatible with the observational data by, e.g., R06. In particular, note the case of Na and Al, for which the yields are predicted to be very sensitive to the metallicity. The IMF (initial mass function) weighted yields, in the [X/Fe] form, increase by about 0.5 and 0.4 dex, respectively, from $\feh\sim-1$ to $\feh\sim0$ (Fig.~5 in K06), yet the data shows no such dependence of [Na/Fe] or [Al/Fe] with [Fe/H] in the thick disk (see Figs.~15 and 16 in R06). Also, in the case of Mn, the observations (Fig.~16 in R06) suggest a strong dependence with metallicity such that the [Mn/Fe] ratio increases by about 0.4 dex from $\feh=-1$ to solar. Our GCE model for Mn predicted an increase of only 0.15~dex in this range (see also Fig.~5 in K06). Oxygen is probably an exceptionally well modeled element because, as stated by K06, it is one of the best described elements in nucleosynthesis.

From Fig.~\ref{f:ofe2}, which separates stars with high kinematic probability of being thin- or thick-disk stars from those with intermediate kinematics, we conclude that the thin/thick disk separation is real, given that including stars with intermediate probabilities does not destroy the bimodal distribution. In other words, stars with kinematics intermediate between that of the thin and thick disks do not belong to another population but are only part of the tails of the kinematic distributions, which clearly overlap. The discussion in Sect.~\ref{s:interm} and Fig.~\ref{f:interm} allows us to reach the same conclusion with more confidence.

Our results are in qualitative agreement with those by B04, who used the forbidden oxygen lines at 630~nm and 636~nm in a selected sample of thin- and thick-disk stars. B04, however, claim to have found thick disk stars in the range $-0.4<\feh<0.2$ whose [O/Fe] decrease towards the solar value. This knee would be the signature of SN Ia in the abundance pattern of thick disk stars (see also Feltzing et~al. 2003) which we, however, cannot confirm.  The fact that we do not see the knee is not due to the characteristics of our sample (see Fig.~\ref{f:bensby} and the discussion in Sect.~\ref{s:ofetrends}). This discrepancy needs to be investigated in more detail.

In view of the age evidence that suggests that thick-disk stars formed before thin-disk stars, it is important to note that if a hiatus of star formation occurred between the formation of the two disks, it would lead to clearly distinct abundance patterns for thin- and thick-disk stars. This, and some of the other arguments presented in this discussion, were already proposed in the early papers by Gratton et~al. (1996) and Fuhrmann (1998). During the period of low or no star formation, no more contribution of SN~II to the ISM is present, leaving AGB stars and possibly SN~Ia (if the duration of the hiatus is long enough) as main contributors of metals to the ISM. This may explain the sudden decrease in [O/Fe] from the thick to the thin disk phase. It does not, however, explain the overlap in the [Fe/H] distributions. Additional information about the formation history of the Galactic disk is thus required.

Thick disk formation scenarios can be divided into top-down and bottom-up models (e.g., Majewski 1993). In the former type of models, the thick disk is an evolutionary stage of the present thin disk, i.e. an intermediate phase between halo and thin disk formation, produced essentially by monolithic collapse, rotation and conservation of the angular momentum. Thick-disk stars are expected to have an age distribution that connects smoothly with that of thin disk stars without an age gap and an abundance gradient with distance from the galactic plane is predicted. Also, in most top-down models, the thin and thick disk abundance patterns are expected to be continuous. This is not consistent with the observational results, including ours. We cannot probe into the existence of a metallicity gradient with distance from the Galactic plane, but that has been rejected by Gilmore et~al. (1995) and Allende~Prieto et~al. (2006).

Some of the bottom-up models, on the other hand, start with an original thin disk that is later heated by different mechanisms. The most accepted scenario in this type of models is heating by satellite accretion. Quinn et~al. (1993) have shown that the accretion of a satellite galaxy into a disk galaxy results in a heated disk which contains stars from both the original thin disk and the accreted satellite. They have also shown that if more than one satellite is accreted, there will be additional heating but only the first accretion event is necessary for thick disk formation. Their simulation does not include hydrodynamical effects so it is not clear what happens to the gas during and after the merger. However, in order to explain the observations, it has been suggested that after the merger, star formation stops and the gas settles down into a new thin disk which is now mixed with the gas from the satellite. If the latter is more metal-poor, we expect the new thin disk to start with an overall metallicity lower than the maximum metallicity reached by the thick-disk stars, thus explaining the overlap in the $\feh$ distributions of thin- and thick-disk stars. In this scenario, due to the hiatus of star formation between thick and thin disk formation, the abundance patterns for thin- and thick-disk stars are expected to be discrete, as explained above. An alternative bottom-up scenario is one in which thick-disk stars are formed in pregalactic fragments before or while they merge and deposit gas and stars into a disk at large scale heights (e.g., Abadi et~al. 2003, Brook et~al. 2005). After the merger(s), the gas settles and cools down, eventually leading to the formation of thin-disk stars.

The details of the different merger scenarios for the formation of the thick-disk have been summarized by R06 (their Sect.~7). In particular, here we emphasize the importance of the recent studies by Brook et~al. (2004, 2005), who use $N$-body simulations that take into account, among other things, gas dynamics, star formation, and metal enrichment due to supernovae and intermediate mass stars, and reasonably reproduce the observational results of R06 as well as ours. Brook et~al.'s models predict the thin/thick disk separation in the [O/Fe] vs. $\feh$ plane (in fact in the more generic [$\alpha$/Fe] vs. $\feh$ plane), extending the thick disk [Fe/H] distribution up to values between$-0.1$ and $-0.5$ depending on the choice of model. In any case, the thick disk abundance trend does not connect with that of the thin disk at high metallicity, i.e., these models do not predict a knee, in good agreement with our results and those by R06.
 
Interestingly, additional Galactic and extragalactic evidence provides even further support to the merger(s) scenario (see Freeman \& Bland-Hawthorn 2002 for a recent review). Abundance patterns for other elements also show, with some exceptions, discrete trends for thin- and thick-disk stars (e.g., Bensby et~al. 2005, R06). Thick disk formation cannot be a natural evolutionary phase of non-interacting galaxies, given that some disk galaxies seem not to have thick disks (Fry et~al. 1999) and thick disks have been shown to be more common in interacting/merging galaxies (Schwarzkopf \& Dettmar 2000). Yoachim \& Dalcanton (2005) have found a galaxy in which the thick disk is rotating in the opposite direction with respect to the thin disk, a situation that cannot be explained by monolithic collapse. In a recent study of thick disks in edge-on galaxies, Yoachim \& Dalcanton (2006) conclude that their observations can be explained by models in which stars form in galactic subunits before merging to form a thick disk. Finally, Wyse et~al. (2006) present an interesting case for streaming stars in the direction of dwarf spheroidal galaxies that could be remnants of the past mergers that led to the formation of the Galactic thick disk.

\bigskip

In summary, the abundance patterns derived in this and previously published works strongly favor a discrete separation between thin- and thick-disk stars. We have demonstrated that an intermediate population with properties between those of the thin and thick disks does not exist. The systematic differences found in the oxygen abundance trends are explained in terms of ISM enrichment by SN~Ia and SN~II, taking into account the importance of each of them at different epochs. In particular, we showed that the oxygen abundances of thick-disk stars are explained by the metallicity dependence of the yields due to massive stars only. Stellar dating indicates that the thick disk was formed prior to the thin disk, while the abundances suggest a hiatus of star formation between thick- and thin-disk formation. With the abundance and age evidence, the observed kinematics of thin/thick disk stars, and observations of thick disks in other galaxies, the merger(s) scenario currently seems to be the most reliable model for the formation of the thick disk. It accounts for a hiatus of star formation between the formation of the two disks and the abundance and kinematic patterns. Finally, the previously proposed existence of a knee in the oxygen abundance pattern of the thick disk has been challenged by our results. We provide some clues to the reason for the discrepancy but a fuller exploration is required, in particular a homogeneous determination of oxygen abundances using different indicators and taking into account all possible sources of systematic error.

\begin{acknowledgements}
This work was supported in part by the Robert~A.~Welch Foundation of Houston, Texas. CAP research is funded by NASA (NAG5-13057 and NAG5-13147). We thank B.~Reddy for sharing spectra from R06 prior to publication, T.~Bensby for sending equivalent widths measured in his observed spectra, K.~Nomoto and C.~Kobayashi for providing their nucleosynthesis yield tables and chemical evolution model results in electronic format, F.~Pont for his assistance with the BAYESAGE code and valuable comments on the determination of stellar ages, and the referee, Raffaele~Gratton, for his suggestions to improve the paper. Some of the data used in this work were obtained at the Hobby-Eberly Telescope (HET), which is a joint project of the University of Texas at Austin, the Pennsylvania State University, Stanford University, Ludwig-Maximilians-Universit\"at M\"unchen, and Georg-August-Universit\"at G\"ottingen. The HET is named in honor of its principal benefactors, William P. Hobby and Robert E. Eberly. We are indebted to D.~Doss as well as the HET and McDonald Observatory staffs for their support during our observations.
\end{acknowledgements}

\begin{appendix}

\section{Systematic errors in our stellar age determination} \label{s:ageerrors}

In R03, AP04, and this work, the stellar ages were determined from the observed stellar parameters and the corresponding values predicted by the theoretical isochrones of Bertelli et~al. (1994). The isochrone points contain the additional information required, namely, stellar ages, and also other important quantities such as the radius and mass. Thus, a comparison of theoretical and observed data provides an estimative of these extra quantities. The actual procedure involves computing probabilities of a given star to have certain ages within the ranges allowed by the observed parameters. The probability distribution is then used to estimate the age and its error (see Appendix~A in AP04a). Here we comment on possible systematic errors associated with this procedure of age determination.

Given the impact of $\alpha$-element enhancement on stellar evolution calculations (e.g., Kim et~al. 2002), the use of models with solar-scaled chemical compositions (like those by Bertelli et~al. 1994) introduces systematic errors in the age determination from isochrones. We compared the ages determined with and without this effect. In addition, since the age determination from isochrones is subject to statistical biases (e.g., Pont \& Eyer 2004, J\o rgensen \& Lindegren 2005), we compared our ages with those derived using the Pont \& Eyer (2004) method, which uses Bayesian statistics and a procedure that attempts to eliminate biases in standard isochrone age determinations. Note also that the Pont \& Eyer method uses the synthetic color-magnitude diagrams computed by Aparicio \& Gallart (2004), which, in addition to the Bertelli et~al. results, is also based on isochrones that consider $\alpha$-element enhancement.

Using the empirical prescription by Degl'Innocenti et~al. (2005), in which the $\alpha$-element enhancement effects on the isochrones are mimicked by increasing the $\feh$ values, as originally proposed by Salaris et~al. (1993),\footnote{Here the original (percentile) metallicity $Z_0$ is increased to $Z=Z_0(af_\alpha+b)$, where $f_\alpha=10^{\mathrm{[\alpha/Fe]}}$ and $a,b$ are coefficients that depend on the adopted solar abundances. The $a,b$ values we used, which are given by Degl'Innocenti et~al. (2005), correspond to the solar abundances by Asplund et~al. (2005).} we find that the ages obtained without $\alpha$-element enhancement are underestimated with respect to those obtained with it. The differences between ages computed with and without the $\alpha$-element enhancement effects are more important for the oldest stars. The differences are as large as 4~Gyr for a 13~Gyr star and 2~Gyr for an 8~Gyr star.

On the other hand, our isochrone ages are in satisfactory agreement with those obtained from the Pont \& Eyer (2004) method, considering the uncertainties of both age estimates. On average, below 6 Gyr, a constant difference of about 1 Gyr is observed, with our mean ages being smaller. From about 10 to 15 Gyr, the differences with the Pont \& Eyer ages increase with age, but this time in the opposite direction (our adopted ages are larger). The largest differences are for the oldest stars, whose ages may be overestimated by about 3~Gyr when using our method. Even though this may be related to the `terminal age bias' discussed by Pont \& Eyer (2004), it is more likely to be explained by systematic differences in the isochrones (note, again, that the input data used by the Pont \& Eyer method includes isochrones computed with $\alpha$-element enhancement).

Tentatively, we conclude that using the Pont \& Eyer (2004) method and/or considering the $\alpha$-element enhancement effect would result in ages substantially lower than those obtained from our method for the oldest stars. Although these studies are aiming at a substantial improvement to the methods of stellar dating, they need to be tested in more detail. Note, for example, that the $\alpha$-element enhancement leads to a thick disk younger than 10~Gyr. Although this is in disagreement with other recent studies of the thick disk (but note that Allende~Prieto et~al. 2006 conclude that thick-disk stars as young as 8~Gyr exist), it solves the problem that, according to our calculations, a few (4) of our stars (and a similar number in the Reddy et~al. studies) result with lower limits in their ages that are greater than the upper limit on stellar ages (the age of the universe) imposed by the \textit{Wilkinson Microwave Anisotropy Probe} (WMAP) measurements ($13.7\pm0.2$~Gyr, Krauss 2003).

It should be noted that these effects have a smaller impact on our $\logg$ determination. For example, if we adopt $\alpha-$element enhanced isochrones our $\logg$ values for dwarf stars, on average, increase only by about 0.05~dex, thus introducing a very small systematic error in the abundances.

\section{A simple GCE model for the thick disk} \label{s:gce}

\begin{table*}
\centering
\begin{tabular}{cccccccccc} \hline\hline
\multirow{2}{*}{$t$ (Gyr)} & \multirow{2}{*}{$\feh$} & \multicolumn{2}{c}{SFR = const.} & \multicolumn{2}{c}{$\propto e^{-t/\mathrm{1 Gyr}}$} & \multicolumn{2}{c}{$\propto e^{-t/\mathrm{4 Gyr}}$} & \multicolumn{2}{c}{Truncated at 1 Gyr} \\
  & & $\ofe$ & [Mn/Fe] & $\ofe$ & [Mn/Fe] & $\ofe$ & [Mn/Fe] & $\ofe$ & [Mn/Fe] \\
\hline
0.05 & -1.66 & 0.601 & -0.536 & 0.601 & -0.536 & 0.601 & -0.536 & 0.601 & -0.536 \\
0.08 & -1.52 & 0.582 & -0.539 & 0.582 & -0.539 & 0.582 & -0.539 & 0.582 & -0.539 \\
0.13 & -1.38 & 0.564 & -0.542 & 0.564 & -0.542 & 0.564 & -0.542 & 0.564 & -0.542 \\
0.20 & -1.24 & 0.544 & -0.545 & 0.545 & -0.544 & 0.545 & -0.545 & 0.544 & -0.545 \\
0.32 & -1.10 & 0.523 & -0.542 & 0.525 & -0.542 & 0.523 & -0.542 & 0.523 & -0.542 \\
0.50 & -0.96 & 0.498 & -0.535 & 0.502 & -0.536 & 0.499 & -0.535 & 0.498 & -0.535 \\
0.79 & -0.82 & 0.473 & -0.518 & 0.482 & -0.523 & 0.475 & -0.519 & 0.473 & -0.518 \\
1.26 & -0.68 & 0.445 & -0.489 & 0.462 & -0.504 & 0.450 & -0.492 & 0.458 & -0.504 \\
2.00 & -0.54 & 0.427 & -0.455 & 0.450 & -0.485 & 0.433 & -0.463 & 0.458 & -0.505 \\
3.16 & -0.40 & 0.415 & -0.417 & 0.442 & -0.470 & 0.421 & -0.431 & 0.454 & -0.502 \\
5.01 & -0.26 & 0.406 & -0.378 & 0.435 & -0.462 & 0.413 & -0.402 & 0.455 & -0.508 \\
\hline
\end{tabular}
\label{t:gce}
\caption{Results of our thick disk GCE model for different choices of the SFR.}
\end{table*}

Simple homogeneous Galactic Chemical Evolution (GCE) models have been described in detail by, e.g., Tinsley (1980), Clayton (1984), Rana (1991), Pagel (1997), and Matteucci (2001). Here we present a very simple GCE model in which several assumptions are made to simplify the problem but remains a reasonably good picture to study the general properties of the chemical evolution of the thick disk. Readers should be warned that this model does not follow the evolution of the gas fraction and does not take into account infall (hence it is not suitable to study the thin/thick disk transition). Our goal with this calculation is to show what the effect of a metallicity dependent yield of SN~II on the evolution of abundance ratios in an isolated interstellar medium (ISM) is. Although it can be extended to include SN Ia yields, we computed the case in which only massive stars contribute to the enrichment of the ISM. Therefore, this model cannot be used to study the G-dwarf problem either. More elaborated models for the thin/thick disk chemical evolution can be found elsewhere (e.g., Chiappini et~al. 1997, Brook et~al. 2005).

We computed the number density of atoms of an element $X$ in the ISM at a time $t$ from the following formula:
\[N_X(t)=\int_0^t\int_{t'_\mathrm{min}}^{t'_\mathrm{max}}\Psi(t'')\Phi[M(t',t'')]Y_X[M(t',t''),Z(t'')]dt''dt'\ ,\]
where $\Psi$ is the star formation rate (SFR), $\Phi$ the initial mass function (IMF), $Y_X$ the yield of $X$ due to supernovae, $M$ the stellar mass and $Z$ the metallicity of the ISM. The limits of integration $t'_\mathrm{min}$ and $t'_\mathrm{max}$ are related to the minimum ($M_\mathrm{min}$) and maximum ($M_\mathrm{max}$) masses of the stars that enrich the ISM in the model: 
\[t'_\mathrm{min}=\max(t'-l_0M_\mathrm{min}^{-2.5},0)\ ,\ \mathrm{and}\]
\[t'_\mathrm{max}=t'-l_0M_\mathrm{max}^{-2.5}\ ,\]
where $l_0$ comes from the relation adopted for the lifetime $\tau$ of a star, $\tau=l_0 M^{-2.5}$, that results from assuming the luminosity of massive stars to be roughly proportional to $M^{3.5}$ (e.g., Demircan \& Kahraman 1991). Clearly, $l_0$ is the lifetime of the Sun (approximately 9 Gyr) and $M$ is given in solar masses. The mass of a star born at $t''$ and exploding at $t'$ is then
\[M(t',t'')=l_0^{0.4}(t'-t'')^{-0.4}\ .\]

We used the mass- and metallicity-dependent yields ($Y_X$) by K06 (see also Nomoto et~al. 2006), who provide yields for normal SN~II as well as hypernovae. We assumed the ratio of hypernovae to normal SN~II to be 0.5, a value that reproduces very well abundance trends in more complex GCE models (K06).

The Salpeter (1955) IMF, $\Phi(M)\propto M^{-2.35}$, was adopted. Then, $\Phi[M(t',t'')]\propto(t'-t'')^{0.94}$, and the integration is straightforward. For the SFR we used $\Psi(t)=\mathrm{const.}$, $\Psi(t)\propto e^{-t/t_0}$, where $t_0=1,2,4$ Gyr, and a SFR truncated at 1~Gyr but the final results were very similar when compared to the observed data. Since in this model we have no control on the total amount of gas we had to assume an age-metallicity relation in order to compute abundance ratios. The following age-metallicity relation:
\[Z(t)=3.56\times10^{-3}t^{0.7}\ ,\]
which is a reasonably good approximation to the theoretical results by Chiappini et~al. (1997), and also to our observed data (see Fig.~\ref{f:ages}), was adopted. We then assumed $\feh=\log(Z/Z_\odot)$. Finally, the abundance ratios were computed as follows:
\[[X/Y]=\log\left(\frac{N_X(t)}{N_Y(t)}\right)-(A_X-A_Y)_\odot\ ,\]
where the difference $(A_X-A_Y)_\odot$ was obtained from the solar abundances. Note that by taking the abundance ratio we eliminate the proportionality constants in the IMF and SFR, which further simplifies the problem. Note also that if the age-metallicity relation is inaccurate, only the time dependence of the ratios would be affected, but not the abundance trends, e.g., the [O/Fe] vs. [Fe/H] relations, as long as the SFR is a slowly varying function of time (in particular for the case of a constant SFR).

The results and implications of this model for the thick disk are discussed in Sect.~\ref{s:discussion} and are shown in Fig.~\ref{f:ofe3} for the case of constant SFR. The [O/Fe] and [Mn/Fe] results are given in Table~\ref{t:gce}.1 for different choices of the SFR.

\end{appendix}

\Online
\scriptsize

\onllongtab{4}{
\begin{longtable}{rrrrrrrrccc}
\caption{\label{t:basicandkin} Basic and kinematic data. \textit{The full version of this table, which will be available online at the CDS, can be obtained from the authors upon request.}}  \\ \hline \hline
HIP & HD & V &	$\pi$	& $V_r$	& $U$ & $V$	&  $W$	& $P_1$	& $P_2$ &	$P_3$ \\
  &  & mag &	mas	& km s$^{-1}$	& km s$^{-1}$ & km s$^{-1}$	&  km s$^{-1}$	&	&  &	 \\	\hline
\endfirsthead
\caption{continued.} \\ \hline \hline
HIP & HD & V &	$\pi$	& $V_r$	& $U$ & $V$	&  $W$	& $P_1$	& $P_2$ &	$P_3$ \\
  &  & mag &	mas	& km s$^{-1}$	& km s$^{-1}$ & km s$^{-1}$	&  km s$^{-1}$	&	&  &	 \\	\hline
\endhead \hline \endfoot
171	&	224930	&	5.75	&	80.6	$\pm$	3.0	&	-36.9	$\pm$	0.4	&	-8.0	$\pm$	0.7	&	-73.1	$\pm$	1.7	&	-31.1	$\pm$	2.0	&	0.23	&	0.76	&	0.01	\\
394	&	225239	&	6.11	&	27.2	$\pm$	4.3	&	4.6	$\pm$	0.3	&	-124.8	$\pm$	19.4	&	-51.3	$\pm$	8.7	&	-10.9	$\pm$	1.5	&	0.27	&	0.71	&	0.02	\\
475	&	70	&	8.22	&	21.0	$\pm$	1.1	&	-28.4	$\pm$	0.2	&	6.8	$\pm$	0.4	&	-31.4	$\pm$	0.4	&	-36.7	$\pm$	2.0	&	0.93	&	0.07	&	0.00	\\
493	&	101	&	7.46	&	26.2	$\pm$	0.8	&	-45.6	$\pm$	0.1	&	45.8	$\pm$	1.1	&	-34.2	$\pm$	0.1	&	17.0	$\pm$	0.5	&	0.95	&	0.05	&	0.00	\\
522	&	142	&	5.71	&	39.0	$\pm$	0.6	&	5.3	$\pm$	0.3	&	-57.4	$\pm$	1.0	&	-37.5	$\pm$	0.6	&	-15.3	$\pm$	0.3	&	0.93	&	0.07	&	0.00	\\
530	&	153	&	8.36	&	8.1	$\pm$	0.9	&	-31.8	$\pm$	0.3	&	-28.2	$\pm$	4.5	&	-48.5	$\pm$	2.4	&	2.8	$\pm$	0.9	&	0.92	&	0.08	&	0.00	\\

$\vdots$ & $\vdots$ & $\vdots$ &	$\vdots$	& $\vdots$	& $\vdots$ & $\vdots$	&  $\vdots$	& $\vdots$	& $\vdots$ &	$\vdots$ \\
\hline
\end{longtable}
}

\onllongtab{5}{
\begin{longtable}{clccclc}
\caption{\label{t:linedata} Line data. The Van der Waals broadening cross section ($\sigma$) for an atom-perturber relative velocity $v_0=10^6$~cm~s$^{-1}$ and velocity parameter $\alpha$ are given. The $\sigma$ values are given in atomic units ($\mathrm{1~a.u.}=2.80\times10^{-17}$~cm$^2$) The damping constants $\gamma_6$ ($FWHM$) per H-atom at $T=10^4$~K and $C_6$, which are used as input data by some spectrum synthesis codes, can be computed from the values given in this table using the following relations (see, e.g., Gray 1992, Anstee \& O'Mara 1995):
\\ $
\frac{\gamma_6}{N_H}=2\left(\frac{4}{\pi}\right)^{\alpha/2}\Gamma\left(\frac{4-\alpha}{2}\right)v\sigma\left(\frac{v}{v_0}\right)^{-\alpha}=17v^{3/5}C_6^{2/5}\ ,
$ \\
where $v=(8kT/\pi\mu)^{1/2}$, $\mu^{-1}=(m_X^{-1}+m_H^{-1})$, $m_X$ is the mass of an atom of the element $X$, $k$ the Boltzmann constant, and $\Gamma$ the gamma function. For the lines marked with a $\dagger$ the damping constants correspond to the modified \"Unsold approximation ($C_6$ multiplied by 2). The last column gives the solar equivalent widths.}  \\ \hline \hline
Wavelength & Species & $EP$	& $\log gf$ & $\sigma$ & $\alpha$ & $EW$ \\
\AA & & eV & & a.u. & & m\AA \\	\hline
\endfirsthead
\caption{continued.} \\ \hline \hline
Wavelength & Species & $EP$	& $\log gf$ & $\sigma$ & $\alpha$ & $EW$ \\
\AA & & eV & & a.u. & & m\AA \\	\hline
\endhead \hline \endfoot
4620.52	&	\ion{Fe}{ii}	&	2.83	&	-3.21	&	185	&	0.306		&	55.6	\\
4629.34	&	\ion{Fe}{ii}	&	2.81	&	-2.28	&	178	&	0.257		&	98.5	\\
4630.12	&	\ion{Fe}{i}	&	2.28	&	-2.52	&	416	&	0.254		&	73.6	\\
4635.85	&	\ion{Fe}{i}	&	2.85	&	-2.34	&	277	&	0.252		&	53.8	\\
4683.56	&	\ion{Fe}{i}	&	2.83	&	-2.41	&	218	&	0.263		&	56.2	\\
4690.14	&	\ion{Fe}{i}	&	3.69	&	-1.61	&	792	&	0.283		&	54.3	\\
4745.80	&	\ion{Fe}{i}	&	3.65	&	-1.27	&	621	&	0.400	$\dagger$	&	78.0	\\
4749.95	&	\ion{Fe}{i}	&	4.56	&	-1.24	&	634	&	0.275		&	35.3	\\
4779.44	&	\ion{Fe}{i}	&	3.42	&	-2.16	&	317	&	0.261		&	40.6	\\
4788.76	&	\ion{Fe}{i}	&	3.24	&	-1.73	&	238	&	0.249		&	66.1	\\
4799.41	&	\ion{Fe}{i}	&	3.64	&	-2.13	&	353	&	0.241		&	33.9	\\
4808.15	&	\ion{Fe}{i}	&	3.25	&	-2.69	&	297	&	0.274		&	27.8	\\
4961.91	&	\ion{Fe}{i}	&	3.63	&	-2.19	&	375	&	0.256		&	25.9	\\
4962.57	&	\ion{Fe}{i}	&	4.18	&	-1.18	&	804	&	0.400	$\dagger$	&	55.3	\\
5044.21	&	\ion{Fe}{i}	&	2.85	&	-2.03	&	713	&	0.238		&	73.5	\\
5054.64	&	\ion{Fe}{i}	&	3.64	&	-1.98	&	353	&	0.313		&	38.3	\\
5145.09	&	\ion{Fe}{i}	&	2.20	&	-3.08	&	351	&	0.272		&	52.6	\\
5187.91	&	\ion{Fe}{i}	&	4.14	&	-1.26	&	724	&	0.400	$\dagger$	&	63.4	\\
5197.58	&	\ion{Fe}{ii}	&	3.23	&	-2.22	&	180	&	0.247		&	81.4	\\
5217.92	&	\ion{Fe}{i}	&	3.64	&	-1.72	&	542	&	0.400	$\dagger$	&	50.7	\\
5228.38	&	\ion{Fe}{i}	&	4.22	&	-1.19	&	809	&	0.278		&	61.2	\\
5234.63	&	\ion{Fe}{ii}	&	3.22	&	-2.18	&	180	&	0.249		&	82.3	\\
5253.46	&	\ion{Fe}{i}	&	3.28	&	-1.57	&	849	&	0.229		&	81.9	\\
5264.81	&	\ion{Fe}{ii}	&	3.23	&	-3.13	&	186	&	0.300		&	40.5	\\
5285.13	&	\ion{Fe}{i}	&	4.43	&	-1.54	&	1046	&	0.282		&	30.6	\\
5288.52	&	\ion{Fe}{i}	&	3.69	&	-1.51	&	353	&	0.297		&	60.0	\\
5293.96	&	\ion{Fe}{i}	&	4.14	&	-1.77	&	701	&	0.400	$\dagger$	&	31.5	\\
5295.31	&	\ion{Fe}{i}	&	4.42	&	-1.59	&	1014	&	0.281		&	28.1	\\
5322.04	&	\ion{Fe}{i}	&	2.28	&	-2.89	&	341	&	0.236		&	60.8	\\
5373.71	&	\ion{Fe}{i}	&	4.47	&	-0.74	&	1044	&	0.282		&	63.8	\\
5379.57	&	\ion{Fe}{i}	&	3.69	&	-1.51	&	363	&	0.249		&	61.8	\\
5386.33	&	\ion{Fe}{i}	&	4.15	&	-1.67	&	930	&	0.278		&	32.7	\\
5414.07	&	\ion{Fe}{ii}	&	3.22	&	-3.58	&	185	&	0.303		&	28.8	\\
5432.95	&	\ion{Fe}{i}	&	4.45	&	-0.94	&	974	&	0.280		&	70.9	\\
5441.34	&	\ion{Fe}{i}	&	4.31	&	-1.63	&	807	&	0.278		&	36.8	\\
5464.28	&	\ion{Fe}{i}	&	4.14	&	-1.58	&	668	&	0.400	$\dagger$	&	38.1	\\
5472.71	&	\ion{Fe}{i}	&	4.21	&	-1.52	&	752	&	0.215		&	45.8	\\
5473.90	&	\ion{Fe}{i}	&	4.15	&	-0.72	&	738	&	0.241		&	84.5	\\
5483.10	&	\ion{Fe}{i}	&	4.15	&	-1.45	&	737	&	0.241		&	47.9	\\
5487.15	&	\ion{Fe}{i}	&	4.42	&	-1.43	&	908	&	0.279		&	41.0	\\
5522.45	&	\ion{Fe}{i}	&	4.21	&	-1.45	&	744	&	0.215		&	43.4	\\
5525.12	&	\ion{Fe}{ii}	&	3.27	&	-3.94	&	184	&	0.280		&	15.1	\\
5525.54	&	\ion{Fe}{i}	&	4.23	&	-1.12	&	749	&	0.238		&	55.8	\\
5543.94	&	\ion{Fe}{i}	&	4.22	&	-1.04	&	742	&	0.238		&	63.5	\\
5546.51	&	\ion{Fe}{i}	&	4.37	&	-1.21	&	825	&	0.278		&	52.9	\\
5560.21	&	\ion{Fe}{i}	&	4.43	&	-1.09	&	895	&	0.278		&	51.9	\\
5584.76	&	\ion{Fe}{i}	&	3.57	&	-2.22	&	296	&	0.271		&	37.9	\\
5618.63	&	\ion{Fe}{i}	&	4.21	&	-1.27	&	732	&	0.214		&	50.5	\\
5619.60	&	\ion{Fe}{i}	&	4.39	&	-1.60	&	808	&	0.277		&	35.7	\\
5633.95	&	\ion{Fe}{i}	&	4.99	&	-0.23	&	635	&	0.270		&	69.9	\\
5635.82	&	\ion{Fe}{i}	&	4.26	&	-1.79	&	928	&	0.279		&	34.0	\\
5638.26	&	\ion{Fe}{i}	&	4.22	&	-0.77	&	730	&	0.235		&	78.2	\\
5650.71	&	\ion{Fe}{i}	&	5.09	&	-0.86	&	1457	&	0.400	$\dagger$	&	42.0	\\
5653.87	&	\ion{Fe}{i}	&	4.39	&	-1.54	&	792	&	0.277		&	38.9	\\
5679.02	&	\ion{Fe}{i}	&	4.65	&	-0.82	&	1106	&	0.291		&	60.0	\\
5701.54	&	\ion{Fe}{i}	&	2.56	&	-2.22	&	361	&	0.237		&	84.8	\\
5705.46	&	\ion{Fe}{i}	&	4.30	&	-1.40	&	744	&	0.231		&	38.8	\\
5717.83	&	\ion{Fe}{i}	&	4.28	&	-1.03	&	758	&	0.209		&	66.6	\\
5731.76	&	\ion{Fe}{i}	&	4.26	&	-1.20	&	727	&	0.232		&	59.4	\\
5775.08	&	\ion{Fe}{i}	&	4.22	&	-1.30	&	648	&	0.400	$\dagger$	&	59.7	\\
5793.92	&	\ion{Fe}{i}	&	4.22	&	-1.60	&	714	&	0.231		&	35.0	\\
5806.73	&	\ion{Fe}{i}	&	4.61	&	-0.95	&	985	&	0.281		&	55.0	\\
5809.22	&	\ion{Fe}{i}	&	3.88	&	-1.74	&	956	&	0.244		&	52.0	\\
5852.22	&	\ion{Fe}{i}	&	4.55	&	-1.23	&	896	&	0.276		&	41.6	\\
5856.09	&	\ion{Fe}{i}	&	4.29	&	-1.46	&	404	&	0.264		&	33.8	\\
5905.67	&	\ion{Fe}{i}	&	4.65	&	-0.69	&	994	&	0.282		&	60.0	\\
5909.97	&	\ion{Fe}{i}	&	3.21	&	-2.60	&	716	&	0.243		&	40.7	\\
5916.25	&	\ion{Fe}{i}	&	2.45	&	-2.99	&	341	&	0.238		&	57.0	\\
5927.79	&	\ion{Fe}{i}	&	4.65	&	-0.99	&	984	&	0.281		&	42.7	\\
5929.68	&	\ion{Fe}{i}	&	4.55	&	-1.31	&	864	&	0.275		&	40.6	\\
5930.18	&	\ion{Fe}{i}	&	4.65	&	-0.17	&	983	&	0.281		&	90.1	\\
5934.65	&	\ion{Fe}{i}	&	3.93	&	-1.07	&	959	&	0.247		&	79.7	\\
5952.72	&	\ion{Fe}{i}	&	3.98	&	-1.34	&	999	&	0.252		&	67.6	\\
6003.01	&	\ion{Fe}{i}	&	3.88	&	-1.06	&	898	&	0.241		&	87.9	\\
6027.05	&	\ion{Fe}{i}	&	4.08	&	-1.09	&	567	&	0.400	$\dagger$	&	65.6	\\
6056.00	&	\ion{Fe}{i}	&	4.73	&	-0.40	&	1029	&	0.286		&	74.8	\\
6079.01	&	\ion{Fe}{i}	&	4.65	&	-1.02	&	920	&	0.276		&	47.9	\\
6082.71	&	\ion{Fe}{i}	&	2.22	&	-3.57	&	306	&	0.271		&	36.2	\\
6085.26	&	\ion{Fe}{i}	&	2.76	&	-3.05	&	344	&	0.260		&	44.0	\\
6093.64	&	\ion{Fe}{i}	&	4.61	&	-1.40	&	866	&	0.274		&	31.7	\\
6096.67	&	\ion{Fe}{i}	&	3.98	&	-1.83	&	963	&	0.250		&	39.0	\\
6127.91	&	\ion{Fe}{i}	&	4.14	&	-1.40	&	574	&	0.400	$\dagger$	&	51.0	\\
6151.62	&	\ion{Fe}{i}	&	2.18	&	-3.30	&	277	&	0.263		&	50.9	\\
6165.36	&	\ion{Fe}{i}	&	4.14	&	-1.46	&	570	&	0.400	$\dagger$	&	46.4	\\
6170.51	&	\ion{Fe}{i}	&	4.80	&	-0.38	&	1057	&	0.289		&	86.1	\\
6173.34	&	\ion{Fe}{i}	&	2.22	&	-2.88	&	281	&	0.266		&	69.7	\\
6180.20	&	\ion{Fe}{i}	&	2.73	&	-2.63	&	335	&	0.259		&	58.4	\\
6187.99	&	\ion{Fe}{i}	&	3.94	&	-1.62	&	903	&	0.244		&	48.9	\\
6200.31	&	\ion{Fe}{i}	&	2.61	&	-2.44	&	350	&	0.235		&	75.7	\\
6213.43	&	\ion{Fe}{i}	&	2.22	&	-2.48	&	280	&	0.265		&	84.7	\\
6229.23	&	\ion{Fe}{i}	&	2.85	&	-2.83	&	350	&	0.250		&	39.4	\\
6232.64	&	\ion{Fe}{i}	&	3.65	&	-1.22	&	446	&	0.400	$\dagger$	&	87.1	\\
6265.13	&	\ion{Fe}{i}	&	2.18	&	-2.55	&	274	&	0.261		&	88.6	\\
6270.23	&	\ion{Fe}{i}	&	2.86	&	-2.54	&	350	&	0.249		&	54.5	\\
6297.79	&	\ion{Fe}{i}	&	2.22	&	-2.74	&	278	&	0.264		&	76.5	\\
6322.69	&	\ion{Fe}{i}	&	2.59	&	-2.43	&	345	&	0.238		&	78.1	\\
6344.15	&	\ion{Fe}{i}	&	2.43	&	-2.92	&	327	&	0.246		&	66.6	\\
6380.74	&	\ion{Fe}{i}	&	4.19	&	-1.32	&	562	&	0.400	$\dagger$	&	54.3	\\
6419.95	&	\ion{Fe}{i}	&	4.73	&	-0.24	&	885	&	0.274		&	87.6	\\
6432.68	&	\ion{Fe}{ii}	&	2.89	&	-3.57	&	169	&	0.204		&	43.9	\\
6481.87	&	\ion{Fe}{i}	&	2.28	&	-2.98	&	308	&	0.243		&	66.6	\\
6518.37	&	\ion{Fe}{i}	&	2.83	&	-2.52	&	336	&	0.251		&	59.9	\\
6593.87	&	\ion{Fe}{i}	&	2.43	&	-2.42	&	321	&	0.247		&	88.0	\\
6597.56	&	\ion{Fe}{i}	&	4.80	&	-0.97	&	893	&	0.276		&	44.7	\\
6609.11	&	\ion{Fe}{i}	&	2.56	&	-2.69	&	335	&	0.245		&	68.6	\\
6703.57	&	\ion{Fe}{i}	&	2.76	&	-3.06	&	320	&	0.264		&	38.5	\\
6750.15	&	\ion{Fe}{i}	&	2.42	&	-2.62	&	335	&	0.241		&	75.8	\\
6752.71	&	\ion{Fe}{i}	&	4.64	&	-1.22	&	778	&	0.274		&	37.6	\\
6810.26	&	\ion{Fe}{i}	&	4.61	&	-1.00	&	873	&	0.275		&	50.4	\\
6828.59	&	\ion{Fe}{i}	&	4.64	&	-0.82	&	768	&	0.277		&	59.1	\\
6841.34	&	\ion{Fe}{i}	&	4.61	&	-0.71	&	759	&	0.267		&	71.9	\\
6842.69	&	\ion{Fe}{i}	&	4.64	&	-1.22	&	896	&	0.279		&	40.7	\\
6843.66	&	\ion{Fe}{i}	&	4.55	&	-0.83	&	736	&	0.216		&	61.7	\\
6855.16	&	\ion{Fe}{i}	&	4.56	&	-0.74	&	637	&	0.400	$\dagger$	&	76.7	\\
6858.15	&	\ion{Fe}{i}	&	4.61	&	-0.94	&	765	&	0.211		&	52.8	\\
6945.21	&	\ion{Fe}{i}	&	2.42	&	-2.48	&	331	&	0.243		&	84.3	\\
6978.85	&	\ion{Fe}{i}	&	2.48	&	-2.50	&	337	&	0.241		&	81.6	\\
6999.88	&	\ion{Fe}{i}	&	4.10	&	-1.46	&	845	&	0.244		&	58.2	\\
7022.95	&	\ion{Fe}{i}	&	4.19	&	-1.15	&	912	&	0.245		&	68.4	\\
7038.22	&	\ion{Fe}{i}	&	4.22	&	-1.20	&	933	&	0.247		&	66.0	\\
7090.38	&	\ion{Fe}{i}	&	4.23	&	-1.11	&	934	&	0.248		&	70.8	\\
7132.99	&	\ion{Fe}{i}	&	4.08	&	-1.65	&	473	&	0.400	$\dagger$	&	44.5	\\
7222.39	&	\ion{Fe}{ii}	&	3.89	&	-3.26	&	187	&	0.301		&	16.3	\\
7224.49	&	\ion{Fe}{ii}	&	3.89	&	-3.20	&	187	&	0.301		&	22.7	\\
7401.69	&	\ion{Fe}{i}	&	4.19	&	-1.55	&	481	&	0.400	$\dagger$	&	41.6	\\
7418.67	&	\ion{Fe}{i}	&	4.14	&	-1.38	&	469	&	0.400	$\dagger$	&	48.8	\\
7449.33	&	\ion{Fe}{ii}	&	3.89	&	-3.27	&	186	&	0.301		&	23.5	\\
7515.83	&	\ion{Fe}{ii}	&	3.90	&	-3.39	&	186	&	0.301		&	13.6	\\
7583.79	&	\ion{Fe}{i}	&	3.02	&	-1.88	&	367	&	0.238		&	85.7	\\
7710.36	&	\ion{Fe}{i}	&	4.22	&	-1.11	&	470	&	0.400	$\dagger$	&	68.6	\\
7711.72	&	\ion{Fe}{ii}	&	3.90	&	-2.50	&	185	&	0.300		&	45.3	\\
7723.21	&	\ion{Fe}{i}	&	2.28	&	-3.62	&	304	&	0.242		&	44.8	\\
7771.94	&	\ion{O}{i}	&	9.15	&	0.37	&	453	&	0.234		&	70.6	\\
7774.17	&	\ion{O}{i}	&	9.15	&	0.22	&	453	&	0.234		&	60.6	\\
7775.39	&	\ion{O}{i}	&	9.15	&	0.00	&	453	&	0.234		&	48.8	\\

\hline
\end{longtable}
}

\onllongtab{6}{
\begin{longtable}{rrlccrrrrrr}
\caption{\label{t:parsandabund} Fundamental parameters, abundances, and ages. \textit{The full version of this table, which will be available online at the CDS, can be obtained from the authors upon request.}}  \\ \hline \hline
HIP	& HD	& $\teff$	& $\logg$	& $v_t$	& $\feh_\mathrm{Fe~I}$	& $\feh_\mathrm{Fe~II}$	& $\feh$ &	$\oxh_\mathrm{LTE}$ &	$\oxh_\mathrm{NLTE}$ &	$\log\mathrm{Age}$ \\
	& 	& K	& cm~s$^{-2}$	& km s$^{-1}$	& dex	& dex	& dex &	dex &	dex &	yr \\	\hline
\endfirsthead
\caption{continued.} \\ \hline \hline
HIP	& HD	& $\teff$	& $\logg$	& $v_t$	& $\feh_\mathrm{Fe~I}$	& $\feh_\mathrm{Fe~II}$	& $\feh$ &	$\oxh_\mathrm{LTE}$ &	$\oxh_\mathrm{NLTE}$ &	$\log\mathrm{Age}$ \\
	& 	& K	& cm~s$^{-2}$	& km s$^{-1}$	& dex	& dex	& dex &	dex &	dex &	yr \\	\hline
\endhead \hline \endfoot
171	&	224930	&	5405	&	4.60	&	0.86	&	-0.86	$\pm$	0.06	&	-0.76	$\pm$	0.06	&	$\cdots$			&	-0.10	$\pm$	0.01	&	-0.16	$\pm$	0.03	&	$\cdots$						\\
394	&	225239	&	5537	&	3.74	&	1.16	&	-0.56	$\pm$	0.04	&	-0.52	$\pm$	0.06	&	-0.51	$\pm$	0.04	&	-0.05	$\pm$	0.03	&	-0.23	$\pm$	0.01	&	$\cdots$						\\
475	&	70	&	5660	&	4.47	&	0.84	&	-0.43	$\pm$	0.04	&	-0.33	$\pm$	0.04	&	-0.35	$\pm$	0.03	&	-0.02	$\pm$	0.01	&	-0.08	$\pm$	0.02	&	$\cdots$						\\
493	&	101	&	5858	&	4.45	&	1.01	&	-0.28	$\pm$	0.04	&	-0.21	$\pm$	0.03	&	-0.21	$\pm$	0.03	&	-0.04	$\pm$	0.08	&	-0.11	$\pm$	0.05	&	$\cdots$						\\
522	&	142	&	6177	&	4.25	&	1.72	&	-0.05	$\pm$	0.08	&	-0.06	$\pm$	0.09	&	-0.02	$\pm$	0.06	&	0.27	$\pm$	0.06	&	0.12	$\pm$	0.04	&	$\cdots$						\\
530	&	153	&	5779	&	3.80	&	1.35	&	-0.11	$\pm$	0.06	&	-0.09	$\pm$	0.04	&	-0.08	$\pm$	0.04	&	0.34	$\pm$	0.04	&	0.14	$\pm$	0.02	&	9.550	$_{-	0.145	}^{+	0.215	}$	\\

$\vdots$	& $\vdots$	& $\vdots$	& $\vdots$	& $\vdots$	& $\vdots$	& $\vdots$	& $\vdots$ &	$\vdots$ &	$\vdots$ &	$\vdots$ \\
\hline
\end{longtable}
}


\begin{thebibliography}{}

\bibitem[]{} Abadi, M. G., Navarro, J. F., Steinmetz, M., \& Eke, V. R.
             2003, \apj, 597, 21

\bibitem[]{} Allende~Prieto, C., Garc\'ia L\'opez, R. J., Lambert, D. L., \& Gustafsson,~B.
             1999, \apj, 527, 879

\bibitem[]{} Allende~Prieto, C., Asplund, M., \& Lambert, D.~L.
             2001, \apj, 556, L63

\bibitem[]{} Allende~Prieto, C., Asplund, M., Garc\'ia L\'opez, R. J., \& Lambert, D.~L.
             2002, \apj, 567, 544

\bibitem[]{} Allende~Prieto, C., Lambert, D. L., Hubeny, I., \& Lanz, T.
             2003a, \apjs, 147, 363

\bibitem[]{} Allende~Prieto, C., Hubeny, I., \& Lambert, D. L.
             2003b, \apj, 591, 1192

\bibitem[]{} Allende~Prieto, C., Barklem, P. S., Lambert, D. L., \& Cunha, K.
             2004a, \aap, 420, 183 (AP04a)

\bibitem[]{} Allende Prieto, C., Asplund, M., \& Fabiani Bendicho, P.
             2004b, A\&A, 423, 1109

\bibitem[]{} Allende Prieto, C., Beers, T. C., Wilhelm, R., et~al.
             2006, \apj, 636, 804 

\bibitem[]{} Anstee, S. D., \& O'Mara, B. J.
             1995, \mnras, 276, 859

\bibitem[]{} Aparicio, A., \& Gallart, C.
             2004, \aj, 128, 1465

\bibitem[]{} Arnett, W. D., Bahcall, J. N., Kirshner, R. P., \& Woosley, S. E
             1989, \araa, 27, 629

\bibitem[]{} Asplund, M., Nordlund, \AA, Trampedach, R., \& Stein, R. F.
             2000, \aap, 359, 743

\bibitem[]{} Asplund, M., \& Garc\'{\i}a P\'erez, A. E.
             2001, \aap, 372, 601

\bibitem[]{} Asplund, M., Grevesse, N., Sauval, J., et~al.
             2004, \aap, 417, 751

\bibitem[]{} Asplund, M., Grevesse, N., Sauval, J.
             2005, in Cosmic Abundances as Records of Stellar Evolution and Nucleosynthesis (ASP), 
             ed. F. N. Bash \& T. G. Barnes, p. 25

\bibitem[]{} Bagnulo, S., Jehin, E., Ledoux, C., et~al.
             2003, The ESO Messenger, 114, 10

\bibitem[]{} Barbier-Brossat, M., \& Figon, P.
             2000, \aaps, 142, 467

\bibitem[]{} Barklem, P. S., Piskunov, N., \& O' Mara, B. J.
             2000, \aaps, 142, 467

\bibitem[]{} Barklem, P. S., \& Aspelund-Johansson, J.
             2005, \aap, 435, 373

\bibitem[]{} Beers, T. C., Drilling, J. S., Rossi, S., et~al.
             2002, \aj, 124, 931

\bibitem[]{} Bensby, T., Feltzing, S., \& Lundstr\"om, I.
             2003, \aap, 410, 527

\bibitem[]{} Bensby, T., Feltzing, S., \& Lundstr\"om, I.
             2004, \aap, 415, 155

\bibitem[]{} Bensby, T., Feltzing, S., Lundstr\"om, I., \& Ilyin, I.
             2005, \aap, 433, 185

\bibitem[]{} Bertelli, G., Bressan, A., Chiosi, C., et~al.
             1994, \aaps, 106, 275

\bibitem[]{} Bi\'emont, E., \& Zeippen, C. J.
             1992, \aap, 265, 850

\bibitem[]{} Blackwell, D. E., Ibbetson, P. A., Petford, A. D., \& Willis, R. B.
             1976, \mnras, 177, 219

\bibitem[]{} Brook, C. B., Kawata, D., Gibson, B. K., \& Freeman, K. C.
             2004, \apj, 612, 894

\bibitem[]{} Brook, C. B., Gibson, B. K., Martel, H., \& Kawata, D.
             2005, \apj, 630, 298

\bibitem[]{} Butler, K., \& Zeippen, C. J.
             1991, J. Phys. IV, Colloq. C1--141

\bibitem[]{} Cabrera-Lavers, A., Garz\'on, F., \& Hammersley, P. L.
             2005, \aap, 433, 173

\bibitem[]{} Casagrande, L., Portinary, L., \& Flynn, C.
             2006, \mnras, 373, 13

\bibitem[]{} Castelli, F., Gratton, R. G., \& Kurucz, R. L.
             1997, \aap, 318, 841

\bibitem[]{} Chen, B., Stoughton, C., Smith, J. A., et~al.
             2001, \apj, 553, 184

\bibitem[]{} Chiappini, C., Matteucci, F., \& Gratton, R.
             1997, \apj, 477, 765

\bibitem[]{} Chiba, M., \& Beers, T. C.
             2000, \aj, 119, 2843

\bibitem[]{} Clayton, D. D.
             1984, \apj, 285, 411

\bibitem[]{} Cowley, C. R.
             1971, The Observatory, 91, 139

\bibitem[]{} Cutri, R. M., Skrutskie M. F., Van Dyk, S., et~al.
             2003, VizieR Online Data Catalog, II/246

\bibitem[]{} Dalcanton, J. J., \& Bernstein, R. A.
             2002, \aj, 124, 1328

\bibitem[]{} Degl'Innocenti, S., Prada Moroni, P. G., \& Ricci, B.
             2005, {\tt [arXiv: astro-ph/0504611]}

\bibitem[]{} Dehnen, W., \& Binney, J. J.
             1998, \mnras, 298, 387

\bibitem[]{} Demircan, O., \& Kahraman, G.
             1991, \apjs, 181, 313

\bibitem[]{} Drawin, H. W.
             1968, Z. Phys., 211, 404

\bibitem[]{} Duflot, M., Figon, P., \& Meyssonnier, N.
             1995, \aaps, 114, 269

\bibitem[]{} Eriksson, K., \& Toft, S. C.
             1979, \aap, 71, 178

\bibitem[]{} ESA 1997,
             The \textit{Hipparcos} and Tycho Catalogues (ESA SP-1200; Noordwijk: ESA)

\bibitem[]{} Feltzing, S., \& Gustafsson, B.
             1998, \aaps, 129, 237

\bibitem[]{} Feltzing, S., Bensby, T., \& Lundstr\"om, I.
             2003, \aap, 397, L1

\bibitem[]{} Freeman, K., \& Bland-Hawthorn, J.
             2002, \araa, 40, 487

\bibitem[]{} Fry, A. M., Morrison, H. L., Harding, P., \& Boroson, T. A.
             1999, \aj, 118, 1209

\bibitem[]{} Fuhrmann, K.
             1998, \aap, 338, 161

\bibitem[]{} Gehren, T., Butler, K., \& Mashonkina, L., et~al.
             2001, \aap, 366, 981

\bibitem[]{} Gilmore, G., \& Reid, N.
             1983, \mnras, 202, 1025

\bibitem[]{} Gilmore, G., Wyse, R. F. G., \& Jones, J. B.
             1995, \aj, 109, 1095

\bibitem[]{} Gratton, R. G., Carreta, E., Matteucci, F., \& Sneden, C.
             1996, in Formation of the Galactic Halo $\ldots$ Inside and Out (ASP),
             ed. H. Morrison \& A. Sarajedini, p. 307

\bibitem[]{} Gratton, R. G., Carreta, E., Eriksson, K., \& Gustafsson, B.
             1999, \aap, 350, 955

\bibitem[]{} Gratton, R. G., Carreta, E., Matteucci, F., \& Sneden, C.
             2000, \aap, 358, 671

\bibitem[]{} Gray, D. F.
             1992, The Observation and Analysis of Stellar Photospheres, 2nd ed.
             (Cambridge University Press), pp. 207 and 217

\bibitem[]{} Gray, D. F., Tycner, C., \& Brown, K.
             2000, \pasp, 112, 328

\bibitem[]{} Greggio, L.
             2005, \aap, 441, 1055

\bibitem[]{} Heiter, U., \& Luck, R. E.
             2005, \aj, 129, 1063

\bibitem[]{} Hibbert, A., Bi\'emont, E., Godefroid, M., \& Vaeck, N.
             1991, J. Phys. B., 24, 3943

\bibitem[]{} Holweger, H., Bard, A., Kock, M., \& Kock, A.
             1991, \aap, 249, 545

\bibitem[]{} Hubeny, I.
             1988, Computer Phys. Comm., 52, 103

\bibitem[]{} Hubeny, I., \& Lanz, T.
             1995, \apj, 439, 875

\bibitem[]{} Iben, I., \& Renzini, A.
             1983, \araa, 21, 271

\bibitem[]{} Johnson, D. R. H., \& Soderblom, D. R.
             1987, \aj, 93, 864

\bibitem[]{} J\o rgensen, B. R., \& Lindegren L.
             2005, \aap, 436, 127

\bibitem[]{} Juri\'c, M., Ivezi\'c, \v{Z}., Brooks, A., et~al.
             2005, {\tt [arXiv: astro-ph/0510520]}

\bibitem[]{} Kim, Y. C., Demarque, P., Yi, S. K., \& Alexander, D. R.
             2002, \apjs, 143, 499

\bibitem[]{} Kiselman, D.
             1993, \aap, 275, 269

\bibitem[]{} Kiselman, D.
             2001, New Astron. Rev., 45, 559

\bibitem[]{} Kobayashi, C., Umeda, H., Nomoto, K., et~al.
             2006, \apj, in press {\tt [arXiv: astro-ph/0608688]}, K06

\bibitem[]{} Korn, A. J., Shi, J., \& Gehren, T.
             2003, \aap, 407, 691

\bibitem[]{} Krauss, L. M.
             2003, \apj, 596, L1

\bibitem[]{} Kurucz, R. L.
             1993, ATLAS9 Stellar Atmosphere Programs and 2 km/s grid.
             Kurucz CD-ROM No. 13. Cambridge, Mass.: Smithsonian Astrophysical Observatory

\bibitem[]{} Kurucz, R. L., Furenlid, I., Brault, J., \& Testerman, L.
             1984, NSO Atlas No.~1: Solar Flux Atlas from 296 to 1300 nm, Sunspot, NSO

\bibitem[]{} Lallement, R., Welsh, B. Y., Vergely, J. L., et~al.
             2003, \aap, 411, 447

\bibitem[]{} Lambert, D. L., Heath, J. E., Lemke, M., \& Drake, J.
             1996, \apjs, 103, 183

\bibitem[]{} Larsen, J. A., \& Humphreys, R. M.
             2003, \aj, 125, 1958

\bibitem[]{} Latham, D. W., Stefanik, R. P., Torres, G., et~al.
             2002, \aj, 124, 1144

\bibitem[]{} Leroy, J. L.
             1999, \aap, 346, 955

\bibitem[]{} Malaroda, S., Levato, H., \& Galliani, S.
             2001, VizieR Online Data Catalog, III/216

\bibitem[]{} Maeder, A.
             1992, \aap, 264, 105

\bibitem[]{} Majewski, S. R.
             1993, \araa, 31, 575

\bibitem[]{} Masana, E., Jordi, C., \& Ribas, I.
             2006, \aap, 450, 735

\bibitem[]{} Matteucci, F.
             2001, The Chemical Evolution of the Galaxy (Kluwer Academic Publishers)

\bibitem[]{} McWilliam, A.
             1997, \araa, 35, 503

\bibitem[]{} Mel\'endez, J.
             2004, \apj, 615, 1042

\bibitem[]{} Mel\'endez, J., Shchukina, N. G., Vasiljeva, I. E., \& Ram\'{\i}rez, I.
             2006, \apj, 642, 1082

\bibitem[]{} Mermilliod, J.-C., Mermilliod, M., \& Hauck, B.
             1997, \aaps, 124, 349

\bibitem[]{} Mishenina, T. V., Soubiran, C., Kovtyukh, V. V., \& Korotin, S. A.
             2004, \aap, 418, 551

\bibitem[]{} Morel, T., \& Micela, G.
             2004, \aap, 423, 677

\bibitem[]{} Nissen, P. E.
             1995, in IAU Symp. 164, Stellar Populations, ed. P. C. van der Kruit \& G. Gilmore
             (Dordrecht: Reidel), 109

\bibitem[]{} Nissen, P. E., Primas, F., Asplund, M., \& Lambert, D. L.               
             2002, \aap, 390, 235

\bibitem[]{} Nomoto, K., Tominaga, N., Umeda, H., et~al.
             2006, Nuclear Physics~A, in press {\tt [arXiv: astro-ph/0605725]}

\bibitem[]{} Nordstr\"om, B., Mayor, M., Andersen, J., et~al.
             2004, \aap, 418, 989

\bibitem[]{} Ojha, D. K.
             2001, \mnras, 322, 426

\bibitem[]{} Pagel, B. E. J.
             1997, Nucleosynthesis and Chemical Evolution of Galaxies (Cambridge University Press)

\bibitem[]{} Pont, F., \& Eyer, L.
             2004, \mnras, 351, 487

\bibitem[]{} Press, W. H., Teukolsky, S. A., Vetterling, W. T., \& Flannery, B. P.
             1992, Numerical Recipes in Fortran, 2nd ed.
             (Cambridge University Press), p. 614

\bibitem[]{} Prochaska, J. X., Naumov, S. O., Carney, B., et~al.
             2000, \aj, 120, 2513

\bibitem[]{} Quinn, P. J., Hernquist, L., \& Fullagar, D. P.
             1993, \apj, 403, 74

\bibitem[]{} Ram\'irez, I., \& Mel\'endez, J.
             2004, \apj, 609, 417

\bibitem[]{} Ram\'irez, I., \& Mel\'endez, J.
             2005a, \apj, 626, 446

\bibitem[]{} Ram\'irez, I., \& Mel\'endez, J.
             2005b, \apj, 626, 465

\bibitem[]{} Ram\'irez, I., Allende~Prieto, C., Redfield, S., \& Lambert, D. L.
             2006, \aap, 459,613

\bibitem[]{} Rana, N. C.
             1991, \araa, 29, 129

\bibitem[]{} Reddy, B. E., Tomkin, J., Lambert, D. L., \& Allende~Prieto, C.
             2003, \mnras, 340, 304 (R03)

\bibitem[]{} Reddy, B. E., Lambert, D. L., \& Allende~Prieto, C.
             2006, \mnras, 367, 1329 (R06)

\bibitem[]{} Reid, N., \& Majewski, S. R.
             1993, \apj, 409, 635

\bibitem[]{} Reyl\'e, C., \& Robin, A. C.
             2001, \aap, 373, 886

\bibitem[]{} Robin, A. C., Reyl\'e, C., Derriere, S., \& Picaud, S.
             2003, \aap, 409, 523

\bibitem[]{} Salaris, M., Chieffi, A., \& Straniero, O.
             1993, \apj, 414, 580

\bibitem[]{} Salpeter, E. E.
             1955, \apj, 121, 161

\bibitem[]{} Santos, N. C., Israelian, G., \& Mayor, M.
             2004, \aap, 415, 1153

\bibitem[]{} Santos, N. C., Israelian, G., Mayor, M., et~al.
             2005, \aap, 437, 1127

\bibitem[]{} Schild, H., \& Maeder, A.
             1985, \aap, 143, L7

\bibitem[]{} Schuler, S. C., Hatzes, A. P., King, J. R., et~al.
             2006, \aj, 131, 1057

\bibitem[]{} Schwarzkopf, U., \& Dettmar, R. J.
             2000, \aap, 361, 451

\bibitem[]{} Shchukina, N., \& Trujillo Bueno, J.
             2001, \apj, 550, 970

\bibitem[]{} Shchukina, N., Trujillo Bueno, J., \& Asplund, M.
             2005, \apj, 618, 939

\bibitem[]{} Siegel, M. H., Majewski, S. R., Reid, I. N., \& Thompson, I. B.
             2002, \apj, 578, 151

\bibitem[]{} Sneden, C.
             1973, PhD Thesis, University of Texas at Austin

\bibitem[]{} Soubiran, C.
             1993, \aap, 274, 181      

\bibitem[]{} Soubiran, C., \& Girard, P.
             2005, \aap, 438, 139

\bibitem[]{} Soubiran, C., Bienaym\'e, O., \& Siebert, A.
             2003, \aap, 398, 141

\bibitem[]{} Steenbock, W., \& Holweger, H.
             1984, \aap, 130, 319

\bibitem[]{} Takeda, Y.
             1994, \pasj, 46, 53

\bibitem[]{} Th\'evenin, F., \& Idiart, T. P.
             1999, \apj, 521, 753

\bibitem[]{} Tinsley, B. M.
             1980, Fund. Cosmic Phys., 5, 287

\bibitem[]{} Tull, R. G., MacQueen, P. J., Sneden, C., \& Lambert, D. L.
             1995, \pasp, 107, 251

\bibitem[]{} Valenti, J. A., \& Fischer, D. A.
             2005, \apjs, 159, 141

\bibitem[]{} Wheeler, J. C., Sneden, C., \& Truran, J. W.
             1989, \araa, 27, 279

\bibitem[]{} Wiese, W. L., Fuhr, J. R., \& Deters, T. M.
             1996, Atomic Transition Probabilities of Carbon, Nitrogen, and Oxygen.
             J.~Phys.~Chem.~Ref.~Data, Monograph No. 7 

\bibitem[]{} Wyse, R. F. G., Gilmore, G., Norris, J. E., et~al.
             2006, \apj, 639, L13

\bibitem[]{} Yoachim, P., \& Dalcanton, J. J.
             2005, \apj, 624, 701

\bibitem[]{} Yoachim, P., \& Dalcanton, J. J.
             2006, \apj, 131, 226

\bibitem[]{} Yong, D., Lambert, D. L., Allende~Prieto, C., \& Paulson, D. B.
             2004, \apj, 603, 697

\end{thebibliography}
\end{document}